\documentclass{article}

\usepackage[letterpaper,top=2cm,bottom=2cm,left=3cm,right=3cm,marginparwidth=1.75cm]{geometry}
\usepackage{enumerate, amsthm, amsmath, amsfonts, thm-restate, xspace}
\usepackage{graphicx}
\usepackage[colorlinks,linkcolor=blue,filecolor=blue,citecolor=blue,urlcolor=blue,pagebackref]{hyperref}

\usepackage[colorinlistoftodos,textsize=tiny,textwidth=2cm,color=red!25!white,obeyFinal]{todonotes}

\title{Facility Location on High-dimensional Euclidean Spaces}

\usepackage[ruled]{algorithm}
\usepackage{algorithmic}

\newcommand{\eps}{\varepsilon}
\newcommand{\card}[1]{\left| #1 \right|}
\newcommand{\norm}[1]{\left|\left| #1 \right|\right|}
\DeclareMathOperator*{\argmax}{arg\,max}
\DeclareMathOperator*{\argmin}{arg\,min}

\newcommand{\alphali}{\alpha_{\mathrm{Li}}}
\newcommand{\calF}{\mathcal{F}}
\newcommand{\calD}{\mathcal{D}}
\newcommand{\calC}{\mathcal{C}}
\newcommand{\R}{\mathbb{R}}

\newtheorem{theorem}{Theorem}[section]
\newtheorem{lemma}[theorem]{Lemma}

\newtheorem{observation}{Observation}

\newtheorem{definition}{Definition}

\numberwithin{theorem}{section}

\begin{document}

\date{}
\title{Facility Location on High-dimensional Euclidean Spaces\thanks{Supported by organization NSF CCF-2236669 and a gift from Google.}}

\author{Euiwoong Lee\thanks{University of Michigan}\and
Kijun Shin\thanks{Seoul National University}}

\maketitle

\begin{abstract}
Recent years have seen great progress in the approximability of fundamental clustering and facility location problems on high-dimensional Euclidean spaces, including $k$-Means and $k$-Median. While they admit strictly better approximation ratios than their general metric versions, their approximation ratios are still higher than the hardness ratios for general metrics, leaving the possibility that the ultimate optimal approximation ratios will be the same between Euclidean and general metrics. Moreover, such an improved algorithm for Euclidean spaces is not known for Uncapaciated Facility Location (UFL), another fundamental problem in the area.

In this paper, we prove that for any $\gamma \geq 1.6774$ there exists $\eps > 0$ such that Euclidean UFL admits a $(\gamma, 1 + 2e^{-\gamma} - \eps)$-bifactor approximation algorithm, improving the result of Byrka and Aardal~\cite{byrka10}. Together with the $(\gamma, 1 + 2e^{-\gamma})$ NP-hardness in general metrics, it shows the first separation between general and Euclidean metrics for the aforementioned basic problems. 
We also present an $(\alphali - \eps)$-(unifactor) approximation algorithm for UFL for some $\eps > 0$ in Euclidean spaces, where $\alphali \approx 1.488$ is the best-known approximation ratio for UFL by Li~\cite{li13}.
\end{abstract}

\section{Introduction}
The (metric) Uncapacitated Facility Location (UFL) is one of the most fundamental problems in computer science and operations research. 
The input of the problem consists of a metric space $(X, d)$, a set of facility locations $\mathcal{F} \subseteq X$, a set of clients $\mathcal{C} \subseteq X$, as well as facility opening costs $\{ f_i \}_{i \in \mathcal{F}}$. The goal is open a subset of centers $S \subseteq \mathcal{F}$ to minimize the sum of the {\em opening cost} $(\sum_{i \in S} f_i)$ and the {\em connection cost} $(\sum_{j \in \mathcal{C}} \min_{i \in S} d(i, j))$.
After intensive research efforts over the years~\cite{guha98, jain02, mahdian02, li13}, the best approximation ratio is $1.488$~\cite{li13} and the best hardness ratio is $1.463$~\cite{guha98}.

As the objective function is the sum of two heterogeneous terms of the opening cost and the connection cost, the natural notion of {\em bifactor approximation} has been actively studied as well. 
Formally, given an instance of UFL, a solution $S \subseteq \calF$ is called an $(\lambda_f, \lambda_c)$-approximation for some $\lambda_f, \lambda_c \geq 1$ if, for any solution $T \subseteq \calF$, the total cost of $S$ is at most $\lambda_f \cdot F^* + \lambda_c \cdot C^*$, where $F^*$, $C^*$ denote the opening and connection cost of $T$ respectively.
In particular, the case $\lambda_f = 1$, also known as a $\lambda_c$-{\em Lagrangian Multiplier Preserving (LMP)} approximation, has been actively studied due to its connection to another fundamental clustering problem of $k$-Median. There is a $(2-\eps)$-LMP approximation for some $\eps > 2 \cdot 10^{-7}$~\cite{cohen2023breaching}, and any $\lambda_c$-LMP approximation for UFL can be translated to $1.307 \cdot \lambda_c$-approximation for $k$-Median~\cite{gowda2023improved}.

Generalizing the hardness of Guha and Khuller~\cite{guha98}, 
Jain, Mahdian and Saberi~\cite{jain02} proved that no $(\lambda_f, \lambda_c)$-approximation polynomial-time algorithm exists for $\lambda_c < 1 + 2e^{-\lambda_f}$ unless $\mathbf{P} = \mathbf{NP}$. (Guha-Khuller's hardness ratio $\gamma \approx 1.463$ is exactly the solution of $\gamma = 1 + 2e^{-\gamma}$.) 
While the optimal value for $\lambda_c$ is not known for small values of $\lambda_f$, Byrka and Aardal gave an algorithm that achieves an $(\lambda_f, 1 + 2e^{-\lambda_f})$-approximation for any $\lambda_f \geq 1.6774$~\cite{byrka10}. 

Euclidean spaces are arguably the most natural metric spaces for facility location and clustering problems. Formally, {\em Euclidean UFL} is a special case of UFL where the underlying metric is $(\R^k, \| . \|_2)$ for some dimension $k$. When $k = O(1)$, this problem admits a PTAS~\cite{cohen2021near}, while the problem remains APX-hard when $k$ is part of the input~\cite{cohen2022johnson}.\footnote{While the cited paper only studies $k$-Median and $k$-Means, the soundness analysis in their Theorem 4.1 (of the arXiv version) can be directly extended to any number of open facilities $k$, implying APX-hardness of Euclidean UFL.}

Recent years have seen active studies on related $k$-Means and $k$-Median on high-dimensional Euclidean spaces~\cite{
ahmadian2019better, 
grandoni2022refined, 
cohen2022improved}, so that the best-known approximation ratios for them are $5.912$ and $2.406$ respectively. While they are strictly lower than the best-known approximation ratios for general metric spaces (which are $9$ and $2.613$), they are still larger than the best-known hardness ratios for general metrics (which are $1+8/e \approx 3.943$ and $1+2/e \approx 1.73$)~\cite{jain02}, which means that it is still plausible that the optimal approximation ratios for $k$-Median and $k$-Means are the same between Euclidean metrics and general metrics.

Our first result is the first strict separation between Euclidean and general metric spaces for UFL. In particular, we show that Euclidean UFL admits a $(1.6774, 1+2e^{-1.6774}-\eps)$ approximation for some universal constant $\eps > 0$, which is NP-hard to do in general metrics.

\begin{restatable}{theorem}{bifactor}
There exists a $(1.6774, 1+2e^{-1.6774}-\eps)$-approximation algorithm for Euclidean UFL for some $\eps \geq 3 \cdot 10^{-42}$. 
\label{thm:main-bifactor}
\end{restatable}

By the result of Mahdian et al.~\cite{mahdian02}, it implies an $(\gamma, 1 + 2e^{-\gamma} - \eps e^{-(\gamma - 1.6774)})$-approximation for any $\gamma \geq 1.6774$. 
Using this result, we are able to slightly improve the approximation ratio for the best-known $(\alphali \approx 1.488)$-unifactor approximation of Li~\cite{li13}. 

\begin{theorem}
There exists a $(\alphali - \eps)$-approximation algorithm for Euclidean UFL for some $\eps \geq 2 \cdot 10^{-45}$. 
\label{thm:main-unifactor}
\end{theorem}

Recent years also have seen great progress on hardness of approximation for clustering problems in high-dimensional Euclidean spaces, including Euclidean $k$-Means and $k$-Median~\cite{cohen2019inapproximability, cohen2022johnson}. We show that similar techniques extend to UFL as well, proving the APX-hardness.

\begin{theorem}
Euclidean UFL is APX-hard.
\end{theorem}

\section{High-level Plan}
\label{sec:preliminaries}
Our work is based on the framework of Byrka and Aardal~\cite{byrka10} who achieved an optimal $(\lambda_f, 1 + 2e^{-\lambda_f})$-bifactor approximation for $\lambda_f \geq 1.6774$ in general metrics. We first review their framework. 
It is based on the following standard linear programming (LP) relaxation:

\begin{align}
\text{Minimize} \quad & \sum\limits_{i \in \mathcal{F}, j \in \mathcal{C}} d(i, j) x_{ij} + \sum\limits_{i \in \mathcal{F}} f_i y_i \notag \\
\text{subject to} \quad & \sum\limits_{i \in \mathcal{F}} x_{ij} = 1 & \forall j \in \mathcal{C}, \notag \\
                        & x_{ij} \leq y_i & \forall i \in \mathcal{F}, j \in \mathcal{C}, \notag \\
                        & x_{ij}, y_i \in [ 0, 1 ] & \forall i \in \mathcal{F}, j \in \mathcal{C}. \notag
\end{align}

The dual formulation is as follows:

\begin{align}
\text{Maximize} \quad & \sum\limits_{j \in \mathcal{C}} v_j \notag \\
\text{subject to} \quad & \sum\limits_{j \in \mathcal{C}} w_{ij} \leq f_i & \forall i \in \mathcal{F}, \notag \\
                        & v_j \leq w_{ij} & \forall i \in \mathcal{F}, j \in \mathcal{C}, \notag \\
                        & w_{ij} \geq 0 & \forall  i \in \mathcal{F}, j \in \mathcal{C}. \notag
\end{align}

A feasible solution $(x, y)$ induces the {\em support graph}, which is defined as the bipartite graph $G = ((\mathcal{F}, \mathcal{C}), E)$ where nodes $i \in \mathcal{F}$ and $j \in \mathcal{C}$ are adjacent iff the corresponding LP variable $x_{ij} > 0$. Two clients $j, j' \in \mathcal{C}$ are considered {\em neighbors} in $G$ if they share the same facility.

Let $(x^*, y^*)$ be a fixed optimal solution to the primal program. The overall cost is divided into the facility cost $F^* = \sum_{i \in \mathcal{F}} f_i y^*_i$ and the connection cost $C^* = \sum_{i \in \mathcal{F}, j \in \mathcal{C}} d(i, j) x^*_{ij}$. Our goal is to round this solution to obtain a solution $S$ whose total cost is at most $\lambda_f F^* + \lambda_c C^*$; it is well known that it implies the $(\lambda_f, \lambda_c)$-approximation defined in the introduction by scaling~\cite{byrka10}, so let us redefine the $(\lambda_f, \lambda_c)$-approximation for the rest of the paper so that $S$ is $(\lambda_f, \lambda_c)$-approximate if its total cost is $\lambda_f F^* + \lambda_c C^*$.

The opening cost and connection cost for individual clients can be further divided using the optimal LP dual solution $(v^*, w^*)$. For each client $j \in \mathcal{C}$, the fractional connection cost is given by $C^*_j = \sum_{i \in \mathcal{F}} d(i, j) x^*_{ij}$, and the fractional facility cost is computed as $F^*_j = v^*_j - C^*_j$. The \textit{irregularity} of the facilities surrounding $j \in \mathcal{C}$ is defined by
\[ r_\gamma(j) = \frac{d(j, \mathcal{D}_j)-d(j, \mathcal{D}_j \cup \mathcal{C}_j)}{F^*_j}. \]
Similarly,
\[ r'_\gamma(j) = (\gamma-1) \cdot r_\gamma(j) = \frac{d(j, \mathcal{D}_j \cup \mathcal{C}_j)-d(j, \mathcal{C}_j)}{F^*_j}. \]
If $\sum_{i \in \mathcal{F}'} y^*_i = 0$, we set $d(j, \mathcal{F}') = 0$. Similarly, when $F^*_j = 0$, we define $r_\gamma(j) = 0$ and $r'_\gamma(j) = 0$. According to the definition, the following conditions hold: the irregularity $0 \leq r_\gamma(j) \leq 1$, the average distance to a close facility $C_j = d(j, \mathcal{C}_j) = C^*_j - r'_\gamma(j) \cdot F^*_j$, and the average distance to a distant facility $D_j = d(j, \mathcal{D}_j) = C^*_j + r_\gamma(j) \cdot F^*_j$. The maximum distance to a close facility is bounded by $M_j \leq D_j$.

\paragraph{Clustering of~\cite{byrka10}.}
At a high level, clustering operates based on the support graph $G = ((\calF, \calC), E)$. For each $c \in \calC$, let $N_c := \{ c' \in \calC : \exists f \in \calF \mbox{ such that } (c, f), (c', f) \in E \}$ be the neighbor of $c$. The clustering algorithm iteratively selects some client $c$ as a {\em cluster center}, put all its neighbors into the cluster, and proceed with the remaining clients. Eventually, all clients are partitioned into one of these clusters. After this, for each cluster, exactly one facility adjacent to cluster center is opened. This ensures that every client is connected to a facility that is not too far from them. Therefore, the criteria for choosing cluster centers and opening facilities will determine the quality of solution.

Starting with a fractional solution of the LP $(x^*, y^*)$ and a parameter $\gamma \in (1, 2)$, 
\cite{byrka10} constructed the {\em facility-augmented solution} 
$(\bar{x}, \bar{y})$, where each $y^*_i$ value is multiplied by $\gamma$ and each client $j \in \calC$ reconfigures its $x^*_{ij}$ values to be fractionally connected to as close facilities as possible. (E.g., $\bar{x}_{ij} > 0$ implies $x^*_{ij} > 0$, but not vice versa.) 
With some postprocessing, one can also assume that $\bar{x}_{ij} \in \{ 0, \bar{y}_i \}$ for every $i \in \calF, j \in \calC$. Then one can categorize every facility near $j \in \mathcal{C}$ into two types: \textit{close} facilities $\mathcal{C}_j = \{ i \in \mathcal{F} \mid \bar{x}_{ij} > 0 \}$ and \textit{distant} facilities $\mathcal{D}_j = \{ i \in \mathcal{F} \mid \bar{x}_{ij} = 0 \text{ and } x^*_{ij} > 0 \}$. This implies that as $\gamma$ increases, the clusters become smaller, and more facilities are opened.

Let the \textit{average distance} from $j \in \mathcal{C}$ to a set of facilities $\mathcal{F}' \subseteq \mathcal{F}$ be defined as $d(j, \mathcal{F}') = \big(\sum\limits_{i \in \mathcal{F}'} d(i, j) y^*_{i}\big)/\big(\sum\limits_{i \in \mathcal{F}'} y^*_i\big)$. 
Then let $C_j := d(j, \calC_j)$, $M_j := \max_{i \in \calC_j} d(j, i)$, and $D_j := d(j, \calD_j)$. We have $C_j \leq M_j \leq D_j$. 

At this point, the support graph is defined by $(\bar{x}, \bar{y})$ solution. Intuitively, we choose the client $j$ with the smallest $C_j + M_j$ as a new cluster center. Given this clustering, the standard randomized rounding procedure is as follows:

\begin{enumerate}
    \item For each cluster center $j$, choose exactly one facility from its neighboring facility set $\{ i \in \calF : (i, j) \in E \}$ according to the $\bar{y}$ values. (Recall that the sum of these values is exactly 1.)
    \item For any facility $i \in \calF$ that is not adjacent to any cluster center in $G$, independently open $i$ with probability $\bar{y}_i$. 
\end{enumerate}
 
\begin{algorithm}
\caption{\textsc{greedy}: \cite{byrka10}'s clustering algorithm}
\begin{algorithmic}
\REQUIRE Support graph \(G = ((\mathcal{F}, \mathcal{C}), E)\)
\STATE $L = \{\}$
\WHILE{\( \mathcal{C} \neq \emptyset \)}
    \STATE $c \gets \argmin\limits_{j \in \mathcal{C}} \, (C_j + M_j)$
\STATE $L \gets L \cup (c, N_c)$
\STATE $\mathcal{C} \gets \mathcal{C} \setminus N_c$
\ENDWHILE
\RETURN $L$
\end{algorithmic}
\label{algo:byrka}
\end{algorithm}

\noindent
Let us consider one client $j \in \calC$ and see how its expected connection cost can be bounded under the above randomized rounding. 
Byrka and Aardal~\cite{byrka10} proved the following properties. 

\begin{itemize}
    \item The probability that at least one facility in $\mathcal{C}_j$ is opened is at least $1-e^{-1}$.
    \item The probability that at least one facility in $\mathcal{C}_j \cup \mathcal{D}_j$ is opened is at least $1-e^{-\gamma}$.
    \item Let client $j' \in \mathcal{C}$ be a neighbor of $j$ in $G$. Then, either $\mathcal{C}_{j'} \setminus (\mathcal{C}_j \cup \mathcal{D}_j) = \emptyset$ or the {\em rerouting cost} $d(j, \mathcal{C}_{j'} \setminus (\mathcal{C}_j \cup \mathcal{D}_j)) \leq d(j', j) + d(j', \mathcal{C}_{j'} \setminus (\mathcal{C}_j \cup \mathcal{D}_j)) \leq D_j + C_{j'} + M_{j'}$ holds. 
    Especially, when $j'$ is the cluster center of $j$, it is at most $C_j + M_j + D_j$. 
    (Li~\cite{li13} refined this bound to $C_j + (3 - \gamma)M_j + (\gamma -1)D_j$.)
\end{itemize}

Then, one can (at least informally) expect that the expected connection cost of $j$ is at most $(1-e^{-1}) C_j + (e^{-1} - e^{-\gamma})D_j + e^{-\gamma}(D_j + C_j + M_j)$. 
It turns out that setting $\gamma \approx 1.6774$ (the solution of $e^{-1} + e^{-\gamma} - (\gamma - 1)(1 - e^{-1} + e^{-\gamma}) = 0$) ensures that this value is at most $(1+2e^{-\gamma})C^*_j$, proving their $(\gamma, 1+2e^{-\gamma})$-bifactor. (See Section~\ref{sec:proof-main} for the formal treatment of their analysis as well as our improvement.)

\paragraph{Exploit the Geometry of Euclidean Spaces.}
In order to strictly improve the approximation ratio, it is natural to attempt to find a cluster $N$ and its center $j'$ where the above inequality holds with some additional slack. Let $cost_{j'}(j) = d(j, \mathcal{C}_{j'} \setminus (\mathcal{C}_j \cup \mathcal{D}_j))$. Intuitively, our goal is to find a cluster $N \subseteq \calC$ with center $j'$ such that 
\begin{equation}
    \sum_{j \in N} cost_{j'}(j) \leq 
    \sum_{j \in N} \big( (1 - \eps_1)C_j + (3 - \gamma)M_j + (\gamma-1)D_j \big).
    \label{eq:target-rerouting-cost}
\end{equation}

The only requirement from the rounding algorithm is that $N_{j'} \subseteq N$. Compared to~\cite{byrka10}'s clustering, we want to shave $\eps_1 \cdot C_j$ on average. 

Let us consider the very special case where $C_j = M_j = D_j = 1$ for every $j \in \calC$; every facility serving $j$ in the original LP solution $(x^*, y^*)$ is at the same distance from $j$. Let $j' \in \calC$ be a cluster center and $j \in \calC$ be in the cluster of $j'$. Then, a simple {\em 3-hop triangle inequality} (just using $\calC_j \cap \calC_{j'} \neq \emptyset$) ensures that $cost_{j'}(j) \leq 3$, and our goal is to improve it to $(3 - \eps_1)$. If $cost_{j'}(j) > 3-\eps_1$, how should the instance look like around $j'$?

\begin{figure}
    \centering
    \includegraphics[scale=0.2]{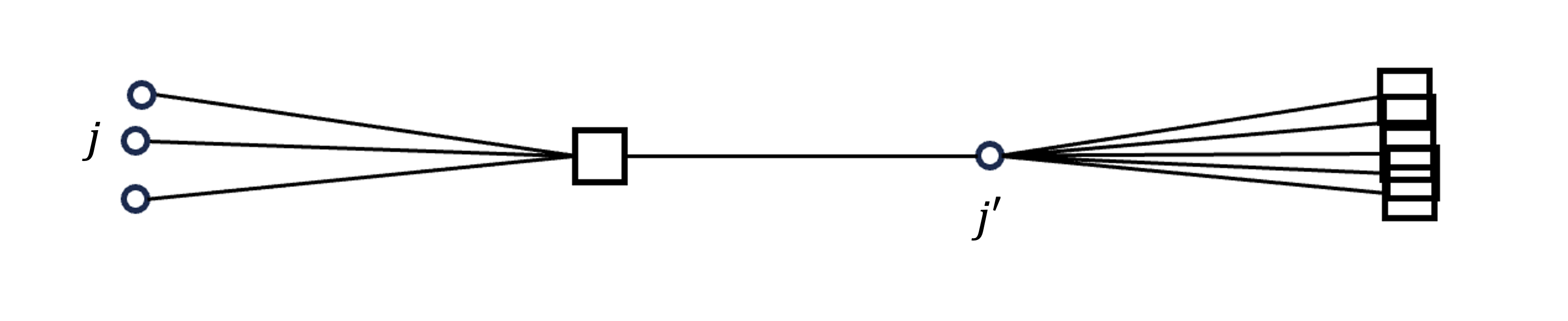}
    \caption{A simple case when $cost_{j'}(j) \approx 3$. 
    There is a client-dense region on the left.}
    \label{fig:colinear}
\end{figure}

It turns out that the instance around $j'$ must exhibit a very specific structure in order to ensure that the 3-hop triangle inequality is tight for almost every neighbor $j \in N_{j'}$. We must have almost every $j \in N_{j'}$ located around almost the same point at distance $2$ from $j'$, where almost all facility neighbors of $j'$ are at the opposite end of the line connecting $N_{j'}$ and $j'$. See Figure~\ref{fig:colinear} for an example. Intuitively, the existence of such a {\em dense region} of clients suggests that if we let a client $j''$ in the region as a new center, many of the 3-hop-triangle inequalities cannot be tight, which implies average rerouting cost $cost_{j''}(j') \leq 3-\eps_1$. If $j''$ is again problematic, we can repeat this procedure over and over.

However, if we relax the condition to $C_j = M_j = 1 \leq D_j$, certain exceptions begin to emerge. One possible scenario is as follows: Since $cost_{j'}(j) = d(j, \mathcal{C}_{j'} \setminus (\mathcal{C}_j \cup \mathcal{D}_j))$ captures the rerouting of $j$ to $j'$'s close facilities $\mathcal{C}_{j'}$ {\em except the $j$'s facilities $\mathcal{C}_j \cup \mathcal{D}_j$}, if $\mathcal{C}_j \cup \mathcal{D}_j$ is large enough to exclude the facilities of $\mathcal{C}_{j'}$, then $cost_{j'}(j)$ might not behave as expected. However, a large volume of $\calC_j \cup \mathcal{D}_j$ implies low $C^*_j / F^*_j$ ratio. If the $C^* / F^*$ ratio is sufficiently low, then this {\em facility-dominant} instance is actually easier to handle with a completely different algorithm, the JMS algorithm~\cite{jain02}, which is known to be $(1.11, 1.7764)$-approximation algorithm.

Therefore, from now on, assume that the cluster centered at $j'$ is connection-dominant. More strictly, assume that for any neighbor $j$ of cluster center $j'$ satisfies that $\calC_j \cup \mathcal{D}_j$ cannot cover half of the ball centered at $j'$ with a radius of $1$. At this point, we can finally assert that it is impossible to avoid the formation of a dense region of clients.

\begin{figure}
    \centering
    \includegraphics[scale=0.21]{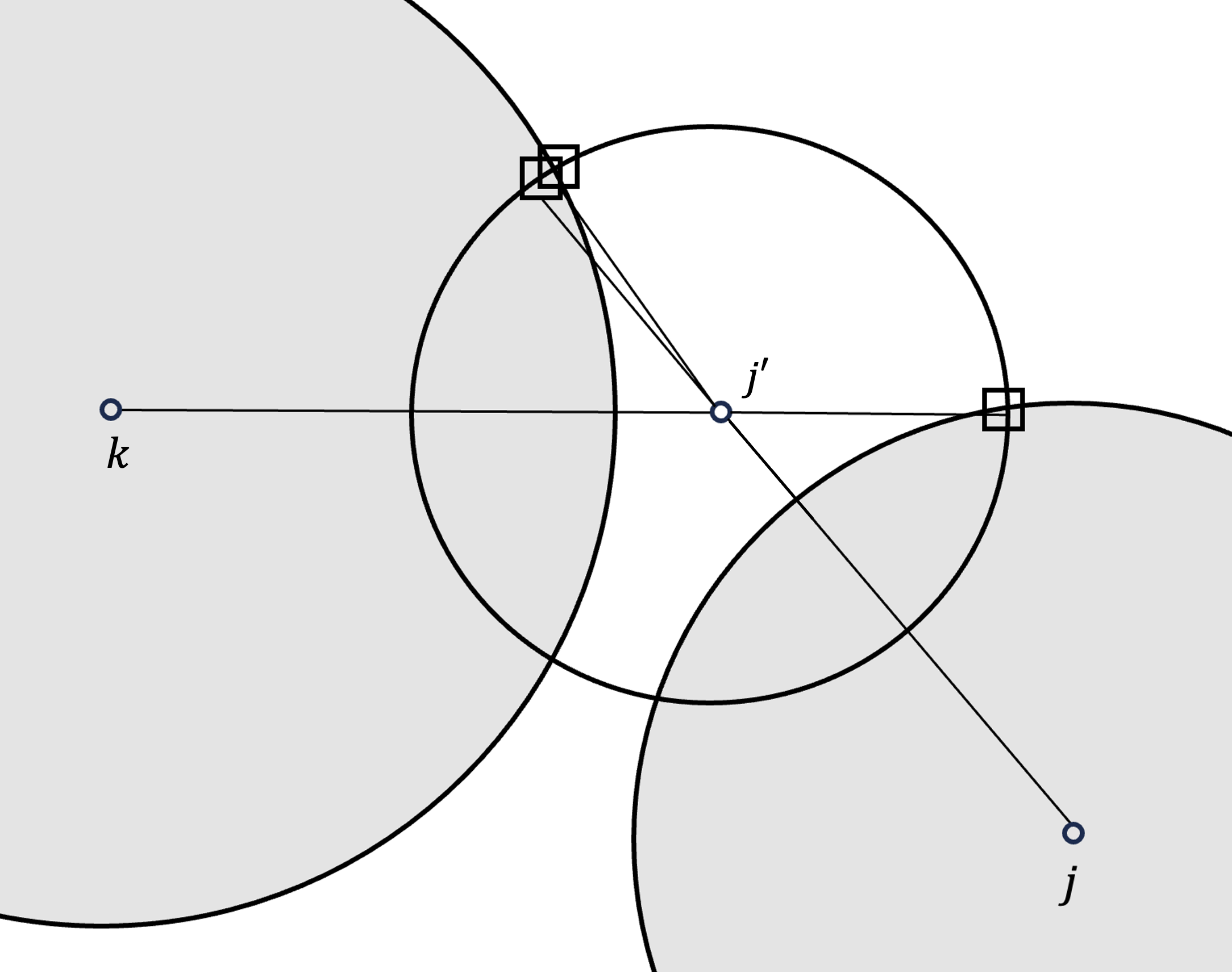}
    \caption{$\calC_{k} \cup \mathcal{D}_{k}$ contains facilities positioned on the opposite side of $j$.}
    \label{fig:perpendicular-visualize}
\end{figure}

Assume towards contradiction that there is no dense region and consider $j$ in $j'$'s cluster such that $cost_{j'}(j) > 3 - \eps_1$. Almost all facilities of $j'$ must be placed in one of two locations: on the opposite side of $j$ relative to $j'$, or within $\calC_j \cup \mathcal{D}_j$. Since there is no dense region, there must be a neighbor $k$ of $j'$ such that $cost_{j'}(k) > 3 - \eps_1$, and $k$ is located in a different direction from $j$. This implies that the facilities positioned on the opposite side of $j$ now help reduce $cost_{j'}(k)$, forcing that they are in $\calC_{k} \cup \mathcal{D}_{k}$; since we assumed that $\calC_{k} \cup \mathcal{D}_{k}$ cannot cover half of the unit ball around $j'$, it implies the angle $\angle jj'k$ must be strictly greater than $\frac{\pi}{2}$!
Ultimately, this process can be simplified to the following situation: inserting unit vectors into a unit sphere with every pairwise angle greater than $\frac{\pi}{2} + \eps$ for some constant $\eps > 0$. It is well known that in the geometry of Euclidean space, there is an upper bound $f(\eps)$ on the number of such vectors, and such an upper bound shows that one of the regions around $j$ (or $k$) we considered must have been dense. See Figure~\ref{fig:perpendicular-visualize} for an example.

However, there are several technical barriers to extending this notion to the general case without restrictions on $C_j$, $M_j$, and $D_j$. In Section~\ref{sec:geom}, we introduce these barriers and formalize the above concept. We also propose sufficient conditions to satisfy~\eqref{eq:target-rerouting-cost}. From Section~\ref{sec:geometric-arguments} to Section~\ref{sec:connection-dominant}, we demonstrate how to remove these conditions, leaving only the connection-dominant instance assumptions. In Section~\ref{sec:proof-main}, we propose and analyze the full algorithm, which achieves improved bi-factor approximation performance.

\section{Finding Good Center via Geometry}
\label{sec:geom}
In this section, we exploit the geometry of Euclidean spaces to prove the existence of a cluster center strictly better than the greedy choice of~\cite{byrka10} under certain conditions (Theorem~\ref{thm:geom-final}). 
We first define several concepts and introduce their motivation, including the sketch of our algorithm.

Recall that our goal is to find a cluster that satisfies~\eqref{eq:target-rerouting-cost}. In all the following propositions, $\gamma$ is a fixed value in the range $\gamma \in (1.6, 2)$.

\begin{definition}
    Suppose $j' \in \mathcal{C}$ is a cluster center. Let $N_{j'}$ be the set of neighbors of $j'$, and $\eps_1 = 10^{-12}$. Additionally, define two more sets:
    \[ N^-_{j'} = \{ j \in N_{j'} \,\,\vert\,\, cost_{j'}(j) > (1-\eps_1)C_j + (3-\gamma)M_j + (\gamma-1)D_j \} \]
    \[ N^+_{j'} = \{ j \in \mathcal{C} \,\,\vert\,\, cost_{j'}(j) \leq (1-\eps_1)C_j + (3-\gamma)M_j + (\gamma-1)D_j \} \]
    Moreover, \textbf{Saving} and \textbf{Spending} of center $j'$ is defined as
    \[ Saving(j') = \sum_{j \in N^+_{j'}} \{ ((1-\eps_1)C_j + (3-\gamma)M_j + (\gamma-1)D_j) - cost_{j'}(j) \}, \]
    \[ Spending(j') = \sum_{j \in N^-_{j'}} \{ cost_{j'}(j) - ((1-\eps_1)C_j + (3-\gamma)M_j + (\gamma-1)D_j) \}. \]
\end{definition}

\noindent

With the goal~\eqref{eq:target-rerouting-cost} in mind, 
$N_{j'}^+$ (resp. $N_{j'}^-$) contains clients $j$ who meet (resp. do not meet) this goal, and $Saving(j')$ (resp. $Spending(j')$) indicates how much $cost_{j'}(j)$ exceeds (resp. is short of) this goal. 
Note that $N^-_{j'} \subseteq N_{j'} \subseteq N^+_{j'} \cup N^-_{j'}$ by definition; it is the {\em best} cluster for center $j'$, which contains all its neighbors (which is required by the algorithm design) and possibly more to increase savings. Therefore, if $Saving(j') \geq Spending(j')$, then $j'$ is considered a \textit{good} center; otherwise, it is a \textit{bad} center.

Now, we will explain some problematic situations that arise when extending the problem to the general case, i.e., without restrictions on $C_j$, $M_j$, and $D_j$'s. The first simple concern is that the choice of the center $j'$ will no longer be solely based on $C_{j'} + M_{j'}$ as in Algorithm~\ref{algo:byrka}, which breaks the previous arguments. Therefore, to gain more flexibility in selecting a new cluster center, it is beneficial to decompose the entire support graph into several layers, where each layer only concerns clients with roughly the same $C_{j'} + M_{j'}$ values. 
\begin{definition}
    A \textbf{network} is a subgraph $((\mathcal{F}', \mathcal{C}'), E')$ of the support graph $G = ((\mathcal{F}, \mathcal{C}), E)$. A network is called \textbf{homogeneous} if there exists $s \geq 0$, such that for any client $j \in \mathcal{C'}$, $s \leq C_j + M_j \leq (1+\delta)s$ for $\delta = 3 \times 10^{-23}$.
\end{definition}

\begin{figure}
    \centering
    \includegraphics[scale=0.2]{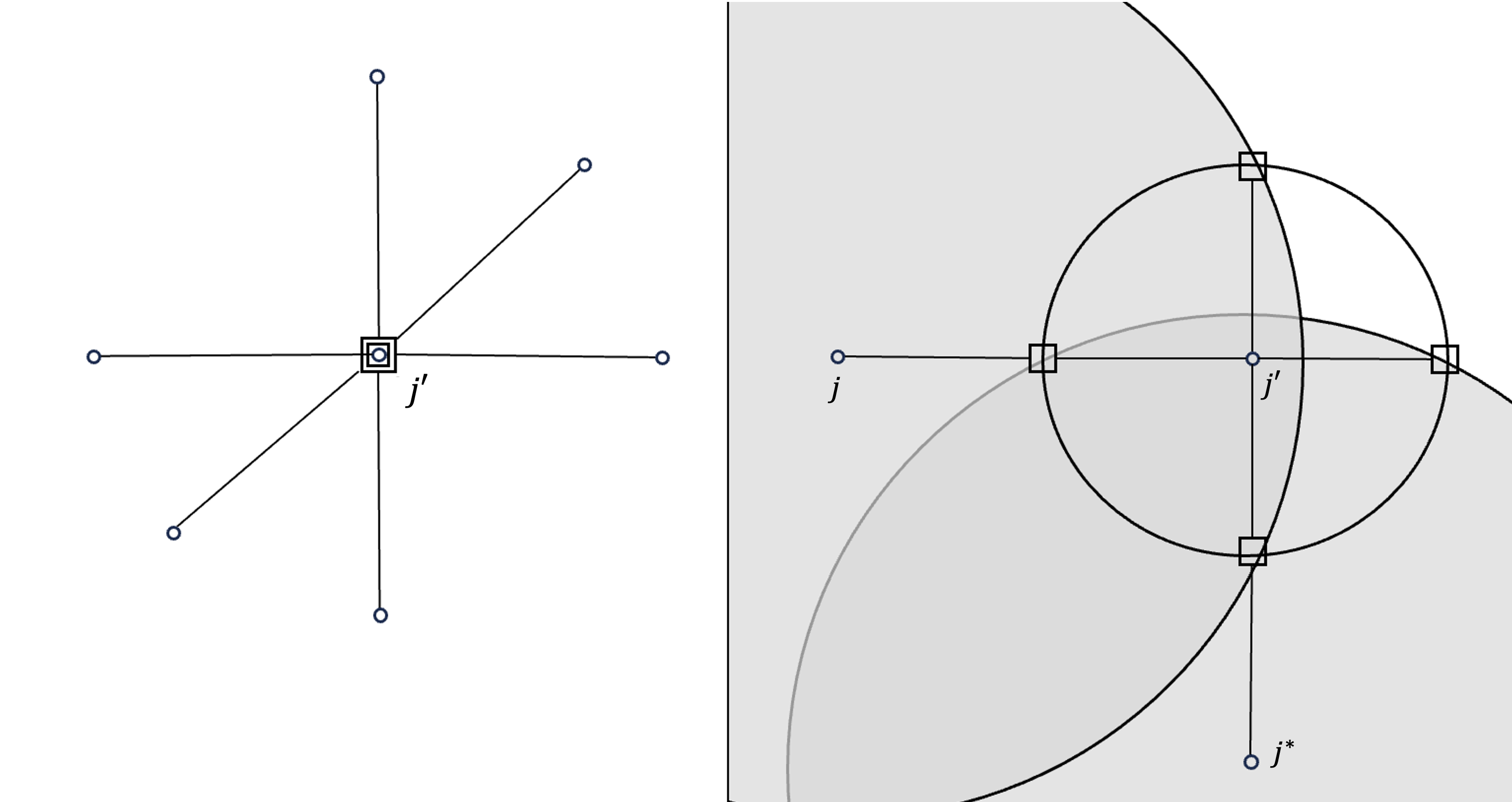}
    \caption{Two examples in a homogeneous network where every neighbor of center $j'$ belongs to $N^-_{j'}$ without creating a dense region.}
    \label{fig:counterexamples}
\end{figure}

\noindent
However, there are still two more scenarios where the above strategy might not hold, as illustrated in Figure~\ref{fig:counterexamples}. This means that the neighbors in $N^-_{j'}$ are all bad clients, but do not create a dense region.

\begin{enumerate}[I.]
\item If $C_{j'} \ll s$, the facilities in $\mathcal{C}_{j'}$ are concentrated near $j'$. In this case, since all the facilities are very close to $j'$, the 3-hop triangle inequality is almost tight for any $j \in N_{j'}$ regardless of where it is.

\item Recall that if $\mathcal{C}_j \cup \mathcal{D}_j$ is large enough to exclude the facilities of $\mathcal{C}_{j'}$, then $cost_{j'}(j)$ might not behave as expected. In particular, a technical problem arises when two facilities from $\mathcal{C}_{j'}$ that are almost antipodal with respect to $j'$ are both contained in $\mathcal{C}_j \cup \mathcal{D}_j$, which is illustrated in Figure~\ref{fig:counterexamples} (right). 
\end{enumerate}

The following definition addresses Scenario I.
\begin{definition}
    For $K_6 = 1.302$, let $\theta = \frac{K_6+1-\gamma}{2K_6+2-\gamma}$. A client $j \in \mathcal{C}$ is \textbf{normal} if $C_{j} \geq \theta(C_{j} + M_{j})$, otherwise \textbf{weird}.
\end{definition}

The following definition addresses Scenario II. 
Note that $v^*_j$ is the maximum distance between $j$ and any facility in $\mathcal{C}_j \cup \mathcal{D}_j$. We are interested in the ball around $j'$ of radius $0.998z_{j'}(j)$ (for sake of analysis), and $j$ having a {\em small remote arm} with respect to $j'$ means that $\mathcal{C}_j \cup \mathcal{D}_j$ cannot contain two antipodal points in our ball of interest (with some slack depending on $\alpha$). Let $z_{j'}(j) = d(j', \mathcal{C}_{j'} \setminus (\mathcal{C}_j \cup \mathcal{D}_j))$. The right-hand side represents the square of the length of the other side of a triangle, where the lengths of the two sides are $d(j', j)$ and $0.998 z_{j'}(j)$, and the included angle between them is $\frac{\pi}{2}-\alpha$.
\begin{definition}
    For normal center $j'$, a neighbor $j$ of $j'$ is said to have a \textbf{small remote arm} if the following holds for $\alpha = 5 \times 10^{-4}$.
    \[ (v^*_j)^2 = (C^*_j+F^*_j)^2 < d(j',j)^2 + \left( 0.998 z_{j'}(j) \right)^2 - 2 d(j',j) \cdot 0.998 z_{j'}(j) \cdot \cos{(\frac{\pi}{2}-\alpha)}
    \]
    Otherwise, $j$ is said to have a \textbf{big remote arm}.
\end{definition}
In Section~\ref{sec:geometric-arguments}, we will show that these two scenarios are the only bad cases to worry about. For instance, if we assume that everything is normal and has a small remote arm, each candidate center $j' \in \calC$ is either good or contains another candidate center $j''$ with $Saving(j'') > Saving(j')$ in its dense region. 

How can we handle these two bad scenarios? In the following two lemmas, we prove that both the weird center and the big remote arm neighbor imply a high ratio of fractional facility cost $F^*_{j'}$ to fractional connection cost $C^*_{j'}$. 

\begin{lemma}
    For any weird client $j$, $C^*_{j} \leq K_6 F^*_{j}$ holds.
    \label{lemma:weird-client}
\end{lemma}
\begin{proof}
     Weird client $j$ satisfies $C_{j} < \theta (C_{j}+M_{j}) \leq \theta (C_{j}+D_{j}) \leq \theta (2C^*_{j} + (2-\gamma)F^*_{j})$, thus
    \[ C^*_{j} < \left( \frac{(\gamma-1)(1-\theta)+\theta}{1-2\theta} \right) F^*_{j} = K_6 F^*_{j}. \]
\end{proof}

\begin{lemma}
    When $1.6 < \gamma < 2$, for a normal center $j'$, if $j \in N^-_{j'}$ has a big remote arm, then $C^*_j \leq K_6 F^*_j$.
    \label{lemma:big-remote-arm}
\end{lemma}
\begin{proof}
    First, $d(j',j) \leq M_{j'} + M_j$ since $j'$ and $j$ are neighbor who share their close facilities. Also $z_{j'}(j) \leq M_{j'}$ according to its definition. Let $C_{j'} = ks$, $z_{j'}(j) = ls$. By the homogeneous condition and normal center condition, 
    \[ \theta s \leq \theta(C_{j'} + M_{j'}) \leq C_{j'} = ks \leq \frac{C_{j'}+M_{j'}}{2} \leq \frac{1+\delta}{2} s, \quad M_{j'} = (C_{j'}+M_{j'}) - C_{j'} \leq (1+\delta-k)s \]
    which implies $\theta \leq k \leq \frac{1+\delta}{2}$ and $l \leq 1-k+\delta$.

    By the triangle inequality and the condition $j \in N^-_{j'}$,
    \begin{align}
        d(j',j) \geq cost_{j'}(j)-z_{j'}(j) &> (1-\eps_1)C_j + (3-\gamma)M_j + (\gamma-1)D_j - M_{j'} \notag \\
        &\geq (C_j+M_j) - (C_{j'}+M_{j'}) + C_{j'} \notag \\
        &\geq -\delta s + ks \geq \frac{2(k-\delta)}{1+\delta} M_{j'} \geq \frac{2(\theta-\delta)}{1+\delta} z_{j'}(j). \notag
    \end{align}
    $\theta$ is minimized at $\gamma = 2$, thus $d(j',j) \geq 0.2319 z_{j'}(j)$. Therefore, by the big remote arm condition,
    \begin{equation}
        (C^*_j+F^*_j)^2 \geq d(j',j)^2 + (0.998 z_{j'}(j))^2 - 2 d(j',j) \cdot 0.998 z_{j'}(j) \cdot \sin{\alpha} \geq 0.995(d(j',j)^2 + z_{j'}(j)^2).
        \label{eq:1}
    \end{equation}
    Since $d(j',j) \geq cost_{j'}(j) - z_{j'}(j)$, by the homogeneous condition, the right hand side can be rewritten as:
    \begin{align}
        d(j',j)^2 + (ls)^2 &> ((1-\eps_1)C_j + (3-\gamma)M_j + (\gamma-1)D_j - ls)^2 + (ls)^2 \notag \\
        &\geq ((3-\gamma-l)s + (\gamma-1)D_j - (2-\gamma+\eps_1)C_j)^2 + l^2 s^2.
        \label{eq:2}
    \end{align}
    Considering the right-hand side as a quadratic function of $s$, it is an increasing function for $s \geq 0$, since $3-\gamma-l > 0$ and $(\gamma-1)D_j \geq (\gamma-1)C_j \geq (2-\gamma+\eps_1)C_j$.
    
    There are two conditions which constrain the lower bound of $s$. At first, the following holds:
    \begin{equation}
        (1+\delta)s \geq C_j+M_j \geq 2C_j.
        \label{eq:3}
    \end{equation}
    Additionally, from $cost_{j'}(j) \leq d(j',j) + z_{j'}(j)$, it follows that
    \[ (1-\eps_1)C_j + (3-\gamma)M_j + (\gamma-1)D_j < M_{j'} + M_j + ls \]
    which implies
    \begin{equation}
        (\gamma-1-\eps_1)(C_j+D_j) < (\gamma-1+l-k+\delta)s
        \label{eq:4}
    \end{equation}
    
    Note that if $F^*_j = 0$, $j$ cannot have a large remote arm. Therefore, by plugging \eqref{eq:3} and \eqref{eq:4} into \eqref{eq:1} bounded by \eqref{eq:2}, we can derive two inequalities. Dividing them by ${F^*_j}^2$ and denoting $x = \frac{C^*_j}{F^*_j}$ and $r_\gamma(j) = r$, they are:
    \begin{align}
        \frac{(x+1)^2}{0.995} &> \left( \left( \frac{6-2\gamma-2l}{1+\delta} - (2-\gamma+\eps_1) \right) (x-(\gamma-1)r) + (\gamma-1)(x+r) \right)^2 \notag \\
        &\quad+ \frac{4l^2}{(1+\delta)^2} (x-(\gamma-1)r))^2, \notag
    \end{align}
    \begin{align}
        &\frac{(x+1)^2}{0.995} - l^2 \left( \frac{(\gamma-1)(2x+(2-\gamma)r)}{\gamma-1+l-k+\delta} \right)^2 \notag \\
        &> \left( \frac{(3-\gamma-l)(\gamma-1)}{\gamma-1+l-k+\delta} (2x+(2-\gamma)r) + (\gamma-1)(x+r) - (2-\gamma+\eps_1)(x-(\gamma-1)r) \right)^2. \notag
    \end{align}
    The intersection of the two derived inequalities offers a range for $x$. Since the range of $\gamma, k, l, r$ is limited, the maximum value of $x$ can be determined through exhaustive search and error propagation, with a detailed process deferred to the Appendix.
\end{proof}

Therefore, if we consider a (sub)instance that has a low facility-to-connection cost ratio, it is natural to expect to apply this argument, which ultimately leads to a (somewhat) good center.
Throughout the paper, we will often express this low facility-to-connection ratio condition will be expressed as $C^* > KF^*$ and use the following values for $K$: $K_1 = 1.3025 > K_2 = 1.3024 > K_3 = 1.3023 > K_4 = 1.3022 > K_5 = 1.3021 > K_6 = 1.302$.
The following theorem is the main result of the section. 

\begin{restatable}{theorem}{geomfinal}
Consider a homogeneous network and 
    let $j'$ be the normal center with the highest saving among all normal centers. Let $c^* = \sum_{j \in N^+_{j'} \cup N^-_{j'}} C^*_j$ and $f^* = \sum_{j \in N^+_{j'} \cup N^-_{j'}} F^*_j$. If $c^* > K_5 f^*$, then this cluster is \textbf{good on average} for $\eps_2 = 5 \times 10^{-18}$ and every $\gamma \in (1.6, 2)$. i.e.
    \[ \sum_{j \in N^+_{j'} \cup N^-_{j'}} cost_{j'}(j) \leq \sum_{j \in N^+_{j'} \cup N^-_{j'}} ((1-\eps_2)C_j + (3-\gamma)M_j + (\gamma-1)D_j). \]
    \label{thm:geom-final}
\end{restatable}

\subsection{Geometric Arguments}
\label{sec:geometric-arguments}
In this subsection, we prove Theorem~\ref{thm:geom-final} using properties of Euclidean geometry.
As previously discussed, our goal is to show: When a bad cluster center $j'$ is normal and (many of) its neighbors have a small remote arm, it is possible to find a dense region of clients near $j'$. 

\begin{figure}
    \centering
    \includegraphics[scale=0.25]{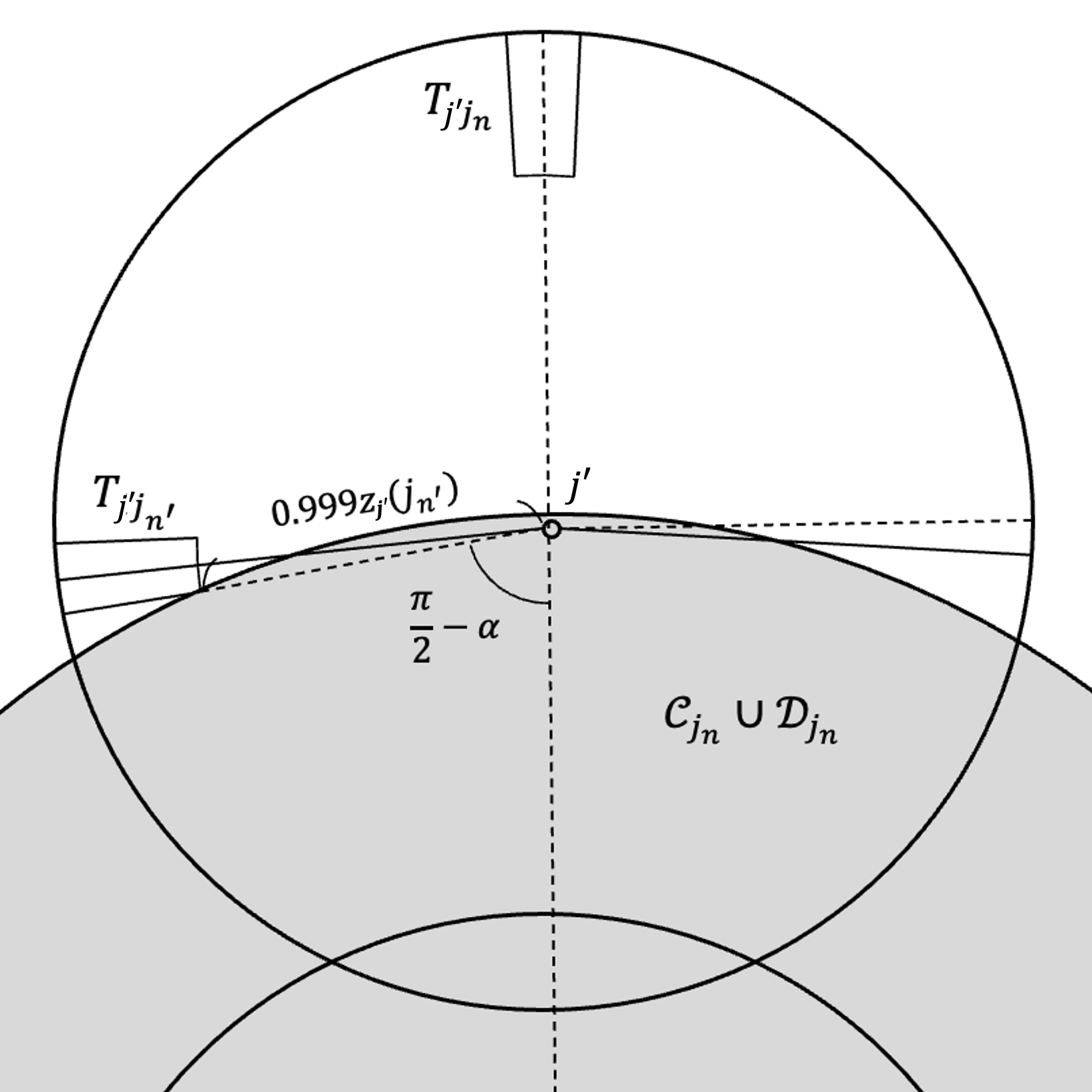}
    \caption{Intersection between a $T_{j'j}$ and a remote arm $\mathcal{C}_j \cup \mathcal{D}_j$.}
    \label{fig:small-remote-arm}
\end{figure}

\noindent When we consider the rerouting of $j$ to $j'$'s close facilities $\mathcal{C}_{j'}$, we can define some {\em worst} facilities. 
In the below definition, $B_{j'j} \subseteq \mathcal{C}_{j'}$ is the set of facilities that have the (almost) worst distance from $j'$, and $T_{j'j} \subseteq B_{j'j}$ is the set of facilities with both worst distance and worst angle. 

\begin{definition}
    Let $r := 10^{-8}$, and let $\phi_r \approx 2 \times 10^{-4}$ be the minimum angle that satisfies
    $1+x^2+2x\cos{\phi_r} \leq (1+(1-r)x)^2$
    for $0 \leq x \leq 2$. For $j \in \mathcal{C}$ and its center $j'$, 
    let 
    \begin{align*}
    B_{j'j} &:= \{ i \in \mathcal{C}_{j'} \,\, \vert \,\, 0.999 z_{j'}(j) \leq d(i, j') \leq M_{j'} \} \\
    T_{j'j} &:= \{ i \in B_{j'j} \,\, \vert \,\, \angle j j' i > \pi - \phi_r \}.
\end{align*}
\end{definition}

The following lemma shows that if $j'$ is normal and $j \in N_{j'}^{-}$ has a bad rerouting through $j'$, then among the facilities in $B_{j'j} \setminus (\mathcal{C}_j \cup \mathcal{D}_j)$, which are the rerouting candidates with worst distance, more than half of them must have a bad angle as well. It is a formalization of the intuition illustrated in Figure~\ref{fig:colinear}. 

\begin{lemma}
    For any $j \in N^-_{j'}$ of normal center $j'$, $z_{j'}(j) > 0.99C_{j'}$.
    \label{lemma:lb-reroute-vector}
\end{lemma}
\begin{proof}
    By the triangle inequality and the homogeneous condition,
    \begin{align}
        z_{j'}(j) \geq cost_{j'}(j)-d(j',j) &> (1-\eps_1)C_j + (3-\gamma)M_j + (\gamma-1)D_j - (M_{j'}+M_j) \notag \\
        &\geq (1-\eps_1)C_j + (2-\gamma)M_j + (\gamma-1)D_j - ((1+\delta)s-C_{j'}) \notag \\
        &\geq -(\eps_1+\delta)s + C_{j'} \geq \left( 1 - \frac{\eps_1 + \delta}{\theta} \right) C_{j'} \geq 0.99C_{j'}. \notag
    \end{align}
\end{proof}

\begin{lemma}
    For a normal center $j'$ and any $j \in N^-_{j'}$, let $G_1 = T_{j'j} \cap (\mathcal{C}_{j'} \setminus (\mathcal{C}_j \cup \mathcal{D}_j))$, $G_2 = (B_{j'j} \setminus T_{j'j}) \cap (\mathcal{C}_{j'} \setminus (\mathcal{C}_j \cup \mathcal{D}_j))$. Then the following holds:
    \[ \sum_{i \in G_1} y^*_i > \sum_{i \in G_2} y^*_i. \]
    \label{lemma:lb-cone-prob}
\end{lemma}
\begin{proof}
    Assume the nontrivial case: $\mathcal{C}_{j'} \setminus (\mathcal{C}_j \cup \mathcal{D}_j) \neq \emptyset$. Let $G_3 = (\mathcal{C}_{j'} \setminus (\mathcal{C}_j \cup \mathcal{D}_j)) \setminus (G_1 \cup G_2)$. Let rerouting probability $p_n$ and rerouting length $l_n$ for $1 \leq n \leq 3$ as:
    \[ p_n = \frac{\sum\limits_{i \in G_n} y^*_i}{\sum\limits_{i \in \mathcal{C}_{j'} \setminus (\mathcal{C}_j \cup \mathcal{D}_j)} y^*_i}, \quad l_n = \frac{\sum\limits_{i \in G_n} y^*_i \cdot d(i, j')}{\sum\limits_{i \in G_n} y^*_i} \]
    Note that $p_1 + p_2 + p_3 = 1$ and $p_1 l_1 + p_2 l_2 + p_3 l_3 = z_{j'}(j)$. Then the goal of this theorem can be written as $p_1 > p_2$. 

    By Lemma \ref{lemma:lb-reroute-vector}, $z_{j'}(j) \geq 0.99C_{j'}$ holds. We derive a lower bound for $p_1+p_2$ as:
    \[ z_{j'}(j) = p_1 l_1 + p_2 l_2 + p_3 l_3 \leq (p_1+p_2) \cdot M_{j'} + (1-p_1-p_2) \cdot 0.999 z_{j'}(j) \]
    implies that
    \[ p_1 + p_2 \geq \frac{0.001z_{j'}(j)}{M_{j'}-0.999z_{j'}(j)} \geq \frac{0.001 \cdot 0.99C_{j'}}{M_{j'}} \geq \frac{0.00099\theta}{1-\theta} \]
    since the right-hand side gives the minimum value when $M_{j'}/C_{j'}$ is the maximum. It is bounded because $j'$ is normal.

    Denote a position vector as $\Vec{v}$. From the definition of $\phi_r$, $cost_{j'}(j)$ is at most
    \begin{align}
        cost_{j'}(j) &= \frac{\sum\limits_{i \in \mathcal{C}_{j'} \setminus (\mathcal{C}_j \cup \mathcal{D}_j)} y^*_i \norm{(\Vec{v_{j'}}-\Vec{v_j}) + (\Vec{v_i}-\Vec{v_{j'}})}}{\sum\limits_{i \in \mathcal{C}_{j'} \setminus (\mathcal{C}_j \cup \mathcal{D}_j)} y^*_i} \notag \\
        &\leq p_1 (d(j',j)+l_1) + p_2 \cdot (d(j',j)+(1-r)l_2) + p_3 (d(j',j)+l_3) \notag \\
        &= d(j',j) + z_{j'}(j) - p_2 r l_2 \notag \\
        &\leq C_{j'} + M_{j'} + (2-\gamma)M_j + (\gamma-1)D_j - p_2 r l_2 \notag \\
        &\leq (1 - 0.999 \cdot 0.99 p_2 r)C_{j'} + M_{j'} + (2-\gamma)M_j + (\gamma-1)D_j. \notag
    \end{align}

    Therefore, since $j \in N^-_{j'}$, $(1-\eps_1)C_j + M_j < (1-0.999 \cdot 0.99 p_2 r)C_{j'} + M_{j'} \leq (1-0.98 p_2 r)C_{j'} + M_{j'}$ holds. It implies
    \begin{align}
        \left( 1-\frac{\eps_1}{2} \right) (C_j+M_j) \leq (1-\eps_1)C_j + M_j &< (1-0.98 p_2 r)C_{j'} + M_{j'} \notag \\
        &\leq ((1-0.98p_2 r) \theta + (1-\theta)) (C_{j'}+M_{j'}). \notag
    \end{align}
    From the homogeneous condition,
    \[ \left( 1-\frac{\eps_1}{2} \right) < (1-0.98 p_2 r\theta)(1+\delta) \]
    which implies
    \[ p_2 < \frac{1}{0.98\theta r} \cdot \frac{\delta + \frac{\eps_1}{2}}{1+\delta} < \frac{0.00099\theta}{1-\theta} < p_1. \]
\end{proof}

The following argument demonstrates how the positional distribution of facilities in $\calC_{j'}$ restricts that of the neighbors in $N^-_{j'}$. Consider two neighbors $j, k \in N^-_{j'}$, both with small remote arms, separated by an angle greater than $2\phi_r$, say $\frac{1}{100}$. Then, $T_{j'j} \cap T_{j'k} = \emptyset$. However, facilities in $T_{j'j}$ reduce $cost_{j'}(k)$ and vice versa, which implies that either $(\calC_{k} \cup \mathcal{D}_{k}) \cap T_{j'j} \neq \emptyset$ or vice versa. 
As the small remote arm condition of $j$ puts a limit on how much $(\calC_{j} \cup \mathcal{D}_{j})$ can intersect $\calC_{j'}$, it is natural to expect that the number of such pairs is small.

\begin{theorem}
    For a normal center $j'$, let $S$ be a subset of $N^-_{j'}$, consisting of clients with a small remote arm. Furthermore, let any two elements $j_1, j_2 \in S$ be separated by an angle greater than $\frac{1}{100}$ with respect to center $j'$, i.e., $\angle j_1 j' j_2 > \frac{1}{100}$. Then, the cardinality of $S$ is bounded by $M = 5 \times 10^6$, independent of the Euclidean space's dimension.
    \label{thm:max-small-remote-arm}
\end{theorem}
\begin{proof}
    Denote $S = \{ j_1, j_2, ..., j_{\card{S}} \}$. Without loss of generality,
    \[ \sum_{i \in T_{j' j_k} \bigcap (\mathcal{C}_{j'} \setminus (\mathcal{C}_{j_k} \cup \mathcal{D}_{j_k}))} y_i^* \geq \sum_{i \in T_{j' j_{k+1}} \bigcap (\mathcal{C}_{j'} \setminus (\mathcal{C}_{j_{k+1}} \cup \mathcal{D}_{j_{k+1}}))} y_i^* \]
    for all $1 \leq k < \card{S}$. Suppose $T_{j' j_{n'}} \cap (\mathcal{C}_{j_n} \cup \mathcal{D}_{j_n}) = \emptyset$ for some $n' < n$. Additionally, given $j_n \in N^-_{j'}$ and the small arm condition, it implies $T_{j' j_n} \cap (\mathcal{C}_{j_n} \cup \mathcal{D}_{j_n}) = \emptyset$. Therefore, $T_{j' j_{n'}} \cup T_{j' j_n} \subseteq \mathcal{C}_{j'} \setminus (\mathcal{C}_{j_n} \cup \mathcal{D}_{j_n})$. Moreover, $T_{j' j_{n'}} \cap T_{j' j_n} = \emptyset$ since $2 \phi_r < \frac{1}{100}$. However, it contradicts to Lemma \ref{lemma:lb-cone-prob}.

    We show that two $j$'s in $S$ with similar $z_{j'}(j)$ values form a large angle with $j'$. 
    Suppose $\angle j_{n'} \, j' \, j_{n} = \beta < \frac{\pi}{2} + \alpha - \phi_r$ for some $n' < n$ with $\frac{z_{j'}(j_n)}{z_{j'}(j_{n'})} \in [\frac{1}{1.001}, 1.001]$.
    Then for any point $x \in T_{j' j_{n'}}$, it holds that $d(j',x) \geq 0.999z_{j'}(j_{n'}) > 0.998z_{j'}(j_n)$ and $\angle j_{n} j' x < \beta + \phi_r < \frac{\pi}{2} + \alpha$. Also, the quadratic function $d(j',x)^2 - 2d(j', j_n) d(j',x) \cdot \sin(\alpha)$ is shown to be non-decreasing for $d(j',x) > 0$. It comes from Lemma \ref{lemma:lb-reroute-vector}, which ensures that $d(j',x) > 0.998z_{j'}(j_n) \geq 0.998 \cdot 0.99 C_{j'} \geq 0.98\theta s \geq 2(1+\delta) \sin{\alpha} \cdot s \geq d(j',j_n) \cdot \sin{\alpha}$. Therefore, since $j_n$ has a small remote arm,
    \begin{align}
        d(j_n, x)^2 &= d(j',j_n)^2 + d(j',x)^2 - 2d(j',j_n)d(j',x) \cdot \cos{\angle j_n j' x} \notag \\
        &> d(j',j_n)^2 + d(j',x)^2 - 2d(j',j_n)d(j',x) \cdot \sin{\alpha} \notag \\
        &\geq d(j',j_n)^2 + (0.998 z_{j'}(j_n))^2 - 2d(j',j_n) \cdot 0.998 z_{j'}(j_n) \cdot \sin{\alpha} \notag \\
        &> (C^*_{j_n}+F^*_{j_n})^2 = {v^*_{j_n}}^2 \notag
    \end{align}
    which implies $T_{j' j_{n'}} \bigcap (\mathcal{C}_{j_n} \cup \mathcal{D}_{j_n}) = \emptyset$, contradicting to the above result. Refer to Figure \ref{fig:small-remote-arm}.
    
    According to Rankin \cite{rankin55}, the maximum number of disjoint spherical caps, each with an angular radius of $\frac{\pi}{4} + \frac{\alpha-\phi_r}{2}$, is at most $1 + \csc(\alpha-\phi_r)$ in any dimension.
    Since $0.99C_{j'} \leq z_{j'}(j) \leq M_{j'}$ holds, it is feasible to segment this range into successive subranges such as $[0.99C_{j'}, 1.001 \cdot 0.99C_{j'}]$, $[1.001 \cdot 0.99C_{j'}, 1.001^2 \cdot 0.99C_{j'}]$, ..., up to $[M_{j'}/1.001, M_{j'}]$. Thus each subrange can only contain a finite number of clients with a small remote arm. Moreover, the normal center condition ensures a bounded number of such divisions. Consequently, the cardinality of $|S|$ is at most
    \[ |S| \leq \left( 1 + \frac{1}{\sin{(\alpha-\phi_r)}} \right) \cdot \frac{\log{\frac{M_{j'}}{0.99C_{j'}}}}{\log{1.001}} \leq \left( 1 + \frac{1}{\sin{(\alpha-\phi_r)}} \right) \cdot \frac{\log{\frac{1-\theta}{0.99\theta}}}{\log{1.001}}. \]
\end{proof}

Therefore, if we have a large $S \subseteq N_{j'}^{-}$ with small remote arms, there must exist a large subset of $S$ whose pairwise angle is small, creating a dense region. 

\begin{lemma}
    For a normal center $j'$, let $S$ be a set of clients from $N^-_{j'}$ with a small remote arm. If $S \neq \emptyset$, then there exists a client $j \in S$ for which $Saving(j) \geq \frac{s}{125} \frac{\card{S}}{M}$.
    \label{lemma:saving-expansion}
\end{lemma}
\begin{proof}
    Let $S'$ be a maximal subset of $S$ where any two clients are separated by an angle greater than $\frac{1}{100}$ with respect to the center $j'$. Then, for any client $j \in S$, there exists a client $z_j \in S'$ such that $\angle j j' z_j \leq \frac{1}{100}$. Therefore, there exists a $z \in S'$ such that the number of clients $j \in S$ for which $z_j = z$ is at least $\frac{\card{S}}{\card{S'}}$.

    For $1 \leq n \leq 5$, let $R_n$ be a region where
    \[ R_n = \{ x \in \mathbb{R}^l \,\, \vert \,\, \angle z \, j' \, x \leq \frac{1}{100}, \,\, \frac{2n-2}{5} (1+\delta)s \leq d(j', x) \leq \frac{2n}{5} (1+\delta)s \}. \]
    Therefore, there exists an index $k$ such that $\card{R_k} \geq \frac{\card{S}}{5\card{S'}}$. For any two clients $j_1, j_2 \in R_k$, the distance $d(j_1, j_2)$ is bounded by the sum of their radial and angular differences. Hence, $d(j_1, j_2) \leq \frac{2(1+\delta)s}{5} + \frac{2(1+\delta)s}{50} = \frac{11}{25} (1+\delta)s$. From the triangle inequality and the homogeneous condition,
    \begin{align}
        cost_{j_1} (j_2) &\leq d(j_1, j_2) + M_{j_1} \leq \frac{36}{25} (1+\delta)s \leq \frac{3-\eps_1}{2(1+\delta)} s \notag \\
        &\leq (1-\eps_1)C_{j_2} + 2M_{j_2} \leq (1-\eps_1)C_{j_2} + (3-\gamma)M_{j_2} + (\gamma-1)D_{j_2} \notag
    \end{align}
    which implies $j_2 \in N^+_{j_1}$. By Theorem \ref{thm:max-small-remote-arm}, $\card{S'} \leq M$. Therefore, the saving of any client $j \in R_k$ is at least
    \begin{align}
        Saving(j) &\geq \frac{\card{S}}{5\card{S'}} \left((1-\eps_1)C_j + (3-\gamma)M_j + (\gamma-1)D_j - \frac{36}{25} (1+\delta)s \right) \notag \\
        &\geq \frac{\card{S}}{5M} \frac{3-72\delta-25\eps_1}{50}s \geq \frac{s}{125} \frac{\card{S}}{M}. \notag
    \end{align}
\end{proof}

Given these geometric tools, 
Theorem~\ref{thm:geom-final} follows as the number of big-remote-arm-neighbors of the chosen center $j'$ can be bounded.

\geomfinal*
\begin{proof}
    Divide $N^+_{j'} \cup N^-_{j'}$ into four groups:
    \begin{enumerate}
        \item Normal clients in $N^-_{j'}$ with a small remote arm,
        \item Weird clients in $N^-_{j'}$ with a small remote arm,
        \item Clients in $N^-_{j'}$ with a big remote arm,
        \item $N^+_{j'}$.
    \end{enumerate}
    
    Let $S_n$ be the set of clients in the $n$-th group ($1 \leq n \leq 4)$. Define the following values:
    \[ C^*_n = \sum_{j \in S_n} C^*_j, F^*_n = \sum_{j \in S_n} F^*_j, spd'_n = \sum_{j \in S_n} \Bigl\{ cost_{j'}(j) - ((1-\eps_2)C_j + (3-\gamma)M_j + (\gamma-1)D_j) \Bigr\}. \]
    
    From the homogeneous condition, $cost_{j'}(j) \leq C_{j'}+M_{j'} + (2-\gamma)M_j + (\gamma-1)D_j \leq (1+\delta)(C_j+M_j) + (2-\gamma)M_j + (\gamma-1)D_j$ holds. Thus,
    \begin{align}
        spd'_n \leq \sum_{j \in S_n} (\delta+\eps_2)(C_j+M_j) \leq \sum_{j \in S_n} (\delta+\eps_2)(C_j+D_j) \leq (\delta+\eps_2)(2C^*_n + (2-\gamma)F^*_n). \notag
    \end{align}
    Summing this for all spending neighbors, then
    \[ \sum_{n=1}^3 spd'_n \leq (\delta+\eps_2)(2c^* + (2-\gamma)f^*) \leq (\delta+\eps_2) \left(\frac{2K_5+2-\gamma}{K_5}\right) c^*. \]

    By Lemma \ref{lemma:weird-client} and Lemma \ref{lemma:big-remote-arm}, $C^*_2 \leq K_6 F^*_2$ and $C^*_3 \leq K_6 F^*_3$ holds. Thus $C^*_2 + C^*_3 \leq K_6(F^*_2 + F^*_3) \leq K_6f^* < \frac{K_6}{K_5}c^*$, which means $C^*_1 + C^*_4 > \left(1-\frac{K_6}{K_5}\right) c^*$. Also $C^*_1+C^*_4 > K_5 (F^*_1+F^*_4)$, since $C^*_2 + C^*_3 \leq K_5(F^*_2 + F^*_3)$. Lastly, from the maximum saving condition, any client $j \in S_1$ satisfies $Saving(j) \leq Saving(j')$. By Lemma \ref{lemma:saving-expansion},
    \[ Saving(j') \geq \frac{s}{125} \frac{\card{S_1}}{M} \geq \frac{\sum_{j \in S_1} (C_j+M_j)}{125M(1+\delta)} \geq \frac{2(C^*_1-(\gamma-1)F^*_1)}{125M(1+\delta)}. \]

    In summary,
    \begin{align}
        \sum_{n=1}^3 spd'_n &\leq (\delta+\eps_2) \left(\frac{2K_5+2-\gamma}{K_5}\right) c^* \leq (\delta+\eps_2) \left(\frac{2K_5+2-\gamma}{K_5-K_6}\right) (C^*_1+C^*_4) \notag \\
        &\leq (\delta+\eps_2) K' ((C^*_1-(\gamma-1)F^*_1) + (C^*_4-(\gamma-1)F^*_4)) \notag \\
        &\leq (\delta+\eps_2) K' \left( \frac{125M(1+\delta)}{2} Saving(j') + \frac{\eps_1-\eps_2}{\eps_1-\eps_2} (C^*_4 - (\gamma-1)F^*_4) \right) \notag \\
        &\leq Saving(j') + (\eps_1-\eps_2)(C^*_4-(\gamma-1)F^*_4) \notag \\
        &\leq \sum_{j \in N^+_{j'}} ((1-\eps_1)C_j + (3-\gamma)M_j + (\gamma-1)D_j - cost_{j'}(j)) + (\eps_1-\eps_2)\sum_{j \in N^+_{j'}} C_j \notag \\
        &= \sum_{j \in N^+_{j'}} ((1-\eps_2)C_j + (3-\gamma)M_j + (\gamma-1)D_j - cost_{j'}(j)) \notag
    \end{align}
    for $K' = \frac{2K_5+2-\gamma}{K_5-K_6} \cdot \frac{K_5}{K_5-\gamma+1}$.
\end{proof}

\section{Clustering for Homogeneous Instances}
\label{section:homo}

In this section, we present an algorithm that operates on a connection-dominant homogeneous instance, ensuring strictly better performance than a naive greedy clustering strategy. Let $c(j)$ for $j \in \calC$ be the center of $j$ when some clustering is given in the context.

\begin{theorem}
    Suppose a homogeneous network $G = ((\mathcal{F}, \mathcal{C}), E)$ satisfies $C^* > K_4 F^*$. Then, the clustering produced by Algorithm \ref{algo:homo} is \textbf{good on average}. Precisely, for $\eps_3 = 3 \times 10^{-32}$ and every $\gamma \in (1.6, 2)$, the following holds:
    \[ \sum_{j \in \mathcal{C}} cost_{c(j)}(j) \leq \sum_{j \in \mathcal{C}} ( (1-\eps_3)C_j + (3-\gamma)M_j + (\gamma-1)D_j ). \]
    \label{thm:homogeneous-clustering}
\end{theorem}

\begin{algorithm}
\caption{\textsc{homogeneous}: Homogeneous Clustering}
\begin{algorithmic}
\REQUIRE Network \(G = ((\mathcal{F}, \mathcal{C}), E)\)
\STATE $L = \{\}$
\WHILE{\( \mathcal{C} \neq \emptyset \)}
    \IF{normal client exists in $\mathcal{C}$}
        \STATE $c \gets \argmax\limits_{j \text{ is normal}} \, Saving(j)$
    \ELSE
        \STATE $c \gets \argmin\limits_{j \in \mathcal{C}} \, (C_j + M_j)$
    \ENDIF
    \STATE $L \gets L \cup (c, N^+_c \cup N^-_c)$
    \STATE $\mathcal{C} \gets \mathcal{C} \setminus (N^+_c \cup N^-_c)$
\ENDWHILE
\RETURN $L$
\end{algorithmic}
\label{algo:homo}
\end{algorithm}

For one of a cluster $A$ made by the algorithm, if the center of $A$ is weird, then the ratio is at most $K_6$ since all clients within them are weird. If $A$ is not `good', its connection-facility ratio is at most $K_5$ by Theorem \ref{thm:geom-final}. Therefore, the assumed ratio $K_4$ in this theorem is larger than these values, implying that a constant proportion of clusters are `good' since they satisfy the conditions of Theorem \ref{thm:geom-final}. Therefore, by reducing the $\eps$ value by that proportion, the desired result can be obtained.

\begin{proof}
    Divide $\mathcal{C}$ into three groups:
    \begin{enumerate}
        \item Clients clustered by a normal center, where the ratio of that cluster's connection cost to facility cost is greater than $K_5$.
        \item Clients clustered by a normal center, where this ratio is at most $K_5$.
        \item Clients clustered by a weird center.
    \end{enumerate}

    Let $S_n$ be the set of clients in the $n$-th group ($1 \leq n \leq 3)$. Here, $S_1$ and $S_2$ correspond to clusters formed by the \textbf{if} condition, and $S_3$ is formed by the \textbf{else} condition. Define the following values:
    \[ C^*_n = \sum_{j \in S_n} C^*_j, \quad F^*_n = \sum_{j \in S_n} F^*_j \]
    
    By Theorem \ref{thm:geom-final}, $\sum_{j \in S_1} cost_{c(j)}(j) \leq \sum_{j \in S_1} ((1-\eps_2)C_j + (3-\gamma)M_j + (\gamma-1)D_j)$. The rerouting cost for $S_2$ is only bounded by the homogeneous condition. For a client $j$, $cost_{c(j)}(j) \leq C_{c(j)}+M_{c(j)} + (2-\gamma)M_j + (\gamma-1)D_j \leq (1+\delta)(C_j+M_j) + (2-\gamma)M_j + (\gamma-1)D_j$. Lastly, clients in $S_3$ are clustered through a greedy strategy. Thus, $cost_{c(j)}(j) \leq C_j + (3-\gamma)M_j + (\gamma-1)D_j$. Note that $S_3$ consists solely of weird clients, meaning $C_3^* \leq K_6 F_3^* \leq K_5 F_3^*$.
    
    From above, the total rerouting cost is bounded by:
    \begin{align}
        \sum_{j \in \mathcal{C}} cost_{c(j)}(j) &= \sum_{j \in S_1} cost_{c(j)}(j) + \sum_{j \in S_2} cost_{c(j)}(j) + \sum_{j \in S_3} cost_{c(j)}(j) \notag \\
        &\leq \sum_{j \in S_1} ((1-\eps_2)C_j+M_j+(2-\gamma)M_j + (\gamma-1)D_j) \notag \\
        &\quad+ \sum_{j \in S_2} ((1+\delta)(C_j+M_j)+(2-\gamma)M_j + (\gamma-1)D_j) \notag \\
        &\quad+ \sum_{j \in S_3} (C_j+(3-\gamma)M_j+(\gamma-1)D_j). \notag
    \end{align}

    Therefore, it is sufficient to show that
    \[ (\delta+\eps_3) \sum_{j \in S_2} (C_j+M_j) + \eps_3 \sum_{j \in S_3} C_j \leq (\eps_2-\eps_3) \sum_{j \in S_1} C_j. \]
    Note that $C^*_2+C^*_3 \leq K_5(F^*_2+F^*_3) < \frac{K_5}{K_4} C^*$ holds, which means $C^*_1 > \left (1-\frac{K_5}{K_4} \right) C^*$. Also $C^*_1 > K_4F^*_1$, since $C^*_2 + C^*_3 \leq K_5 (F^*_2 + F^*_3)$. Then the following holds:
    \begin{align}
    &\quad (\delta+\eps_3) \sum_{j \in S_2} (C_j+M_j) + \eps_3 \sum_{j \in S_3} C_j \notag \\
    &\leq (\delta+\eps_3) \sum_{j \in S_2 \cup S_3} (C_j+M_j) \leq (\delta+\eps_3) \sum_{j \in \mathcal{C}} (C_j + D_j) \notag \\
    &\leq (\delta+\eps_3)(2C^* + (2-\gamma)F^*) \notag \\
    &\leq \frac{(\delta + \eps_3)(2-\gamma+2K_4)}{K_4} C^* \notag \\
    &\leq (\eps_2-\eps_3) \frac{(K_5-\gamma+1)(K_4-K_5)}{K_4 K_5} C^* \leq (\eps_2-\eps_3) \frac{K_5-\gamma+1}{K_5} C^*_1 \notag \\
    &\leq (\eps_2-\eps_3)(C^*_1 - (\gamma-1)F^*_1) \leq (\eps_2-\eps_3) \sum_{j \in S_1} C_j. \notag
    \end{align}
\end{proof}

\section{Clustering for Connection-dominant Instances}
\label{sec:connection-dominant}
In this section, we introduce an algorithm that operates on connection-dominant instances without a homogeneous condition. This algorithm uses Algorithm \ref{algo:byrka} and Algorithm \ref{algo:homo} as subroutines. The theorem stated below shows our main result.

\begin{theorem}
    For any connection-dominant instance, i.e. $C^* > K_1F^*$, there exists an algorithm that finds a clustering configuration whose rerouting cost is at most
    \[ \sum_{j \in \mathcal{C}} cost_{c(j)}(j) \leq \sum_{j \in \mathcal{C}} ((1-\eps_5)C_j + (3-\gamma)M_j + (\gamma-1)D_j) \]
    for $\eps_5 = 2 \times 10^{-41}$ and every $\gamma \in (1.6, 2)$.
    \label{thm:new-algo-perf}
\end{theorem}

\begin{definition}
    Let $B_0$ be the set of clients for which $C_j+M_j = 0$. Let $s = \min\limits_{j \in \mathcal{C} \setminus B_0} (C_j+M_j)$. For $\delta' = 7 \times 10^{-32}$, let a \textbf{Block} $B_n$ be the set of clients such that
    \[ B_n = \{ j \in \mathcal{C} \,\vert\, (1+\delta')^{n-1} s \leq C_j+M_j < (1+\delta')^{n} s \} \]
\end{definition}

Note that if $(1+\delta')^m \leq (1+\delta)$, then a network composed of at most $m$ consecutive blocks is still homogeneous. However, applying Algorithm \ref{algo:homo} directly to each block would be impossible when the block is a facility-dominant or neighbors of the center from some block may belong to a different block. Moreover, the following fact implies that the neighbor relationships between consecutive blocks are the main point.
\begin{observation}
    If two neighbors $j, j' \in \mathcal{C}$ belonging to neither the same block nor consecutive blocks, satisfying $(1+\delta')(C_j+M_j) \leq C_{j'} + M_{j'}$, then $j' \in N^+_j$ holds when $Saving( \cdot )$ criteria $\eps$ satisfies that $\eps \leq \frac{2\delta'}{1+\delta'}$.
    \label{obs:far-block}
\end{observation}
\begin{proof}
\begin{align}
    cost_j(j') &\leq C_j+M_j + (2-\gamma)M_{j'}+(\gamma-1)D_{j'} \leq \frac{C_{j'}+M_{j'}}{1+\delta'} + (2-\gamma)M_{j'}+(\gamma-1)D_{j'} \notag \\
    &\leq \left( 1-\frac{2\delta'}{1+\delta'} \right) \cdot C_{j'} + (3-\gamma)M_{j'} + (\gamma-1)D_{j'}. \notag
\end{align}
\end{proof}

A key idea is that with a sufficiently small $\delta'$, the support graph can be segmented almost arbitrarily while still preserving the homogeneous condition, enabling the identification of weak \textit{connections} between consecutive blocks. Notably, even if two clients $j \in B_n$ and $j' \in B_{n+1}$ come from consecutive blocks, the $cost_j(j')$ is not worse than the greedy strategy. 

\begin{definition}
    An \textbf{interval} $I$ is the set of consecutive blocks, up to a maximum $2L$ for $L = 2 \times 10^8$. Precisely, $I = \{ B_i, \ldots, B_{i+k} \}$ for $k < L$, where $\sum_{j=i}^{i+k} C^*(B_j) - C^*(B_i) \geq K_3 \sum_{j=i}^{i+k} F^*(B_j)$, except when $k = 0$. The \textbf{size} of an interval $I$ is the number of blocks it contains, denoted as $\card{I}$. The \textbf{reward} of interval $I$ is defined as $R(I) = \sum_{j=i}^{i+k} C^*(B_j) - C^*(B_i)$.
\end{definition}

Note that a block is a unique type of interval with a size of $1$. Also $(1+\delta')^{2L} \leq 1+\delta$ holds. An interval is the basic unit of clustering to which Algorithm \ref{algo:byrka} and Algorithm \ref{algo:homo} are applied. Therefore, the reward of an interval represents the extent to which a guaranteed $Saving$ can be obtained when the current interval can be clustered using Algorithm \ref{algo:homo}, regardless of how the preceding intervals have been clustered. From this perspective, the first block, which does not contribute to the reward, serves as a kind of `buffer'. 

Given the entire support graph $G = (\calF \cup \calC, E)$ and a subset $\calC' \subseteq \calC$, let $G_{\calC'}$ be the subgraph (network) induced by $\calF \cup \calC'$.  
Consequently, when we call Algorithms~\ref{algo:byrka} or~\ref{algo:homo} with $G_{\calC'}$, the calculations for $Saving$ and $Spending$ are performed solely with respect to the implicitly defined client set $\calC'$. We denote Algorithm \ref{algo:byrka} as \textsc{greedy} and Algorithm \ref{algo:homo} as \textsc{homogeneous}.
For simplicity, when $I$ denotes an interval, we interpret the expression $\sum_{j \in I}$ as $\sum_{B \in I} \sum_{j \in B}$, aggregating over all clients within the interval. 
The following theorem illustrates how reward is related to $Saving$:
\begin{lemma}
    Let $J$ be a set of non-overlapping intervals. Then clustering produced by Algorithm~\ref{algo:conn} is \textbf{good on average}. Precisely, for $\eps_4 = 2 \times 10^{-36}$,
    \[ \sum_{j \in \mathcal{C}} cost_{c(j)}(j) \leq \sum_{j \in \mathcal{C}} \left( \left( 1-\frac{\sum\limits_{I \in J} R(I)}{\sum\limits_{j' \in \mathcal{C}} C_{j'}} \eps_4 \right) C_j + (3-\gamma)M_j + (\gamma-1)D_j \right). \]
    \label{lemma:a2-perf}
\end{lemma}
\begin{algorithm}
\caption{\textsc{conn}: Connection-dominant Clustering}
\begin{algorithmic}
\REQUIRE Set of disjoint intervals \(J = \{ I_1, I_2, \ldots, I_m \} \), arranged in the increasing order of $C_j + M_j$ values of their clients.
\STATE $L = \{\}$
\FOR{$k=1$ to $m$}
    \IF{$\card{I_k} \geq 2$ and $\sum_{j \in I_k} C^*_j > K_4 \sum_{j \in I_k} F^*_j$}
        \STATE $L \gets L \cup \textsc{homogenous}(G_{I_k})$
    \ELSE
        \STATE $L \gets L \cup \textsc{greedy}(G_{I_k})$
    \ENDIF
    \STATE $S := \bigcup_{j \in I_k} \left( N^+_j \cup N^-_j \right)$
    \FOR{$l=k$ to $m$}
        \STATE $I_l \gets I_l \setminus S$
    \ENDFOR
\ENDFOR
\RETURN $L$
\end{algorithmic}
\label{algo:conn}
\end{algorithm}

\begin{proof}
    Let $\eps' = \min(\eps_3, \frac{2\delta'}{1+\delta'})$. As previously discussed, a client has a low rerouting cost, specifically $cost_{c(j)}(j) \leq (1-\eps')C_j + (3-\gamma)M_j + (\gamma-1)D_j$, in two scenarios: firstly, if the client is clustered by the Algorithm \ref{algo:homo}, which is guaranteed by Theorem \ref{thm:homogeneous-clustering}—noting that this does not bound individual rerouting costs but rather the average cost; secondly, if $c(j)$ is in a block that precedes the block containing $j$ by at least two others, as per Observation \ref{obs:far-block}. Denote the set of such clients by $V$.
    
    An interval $I \in J$ can be clustered in two ways:
    \begin{enumerate}
        \item If interval $I$ is selected via an if condition, then any client $j \in I$, not in $I$'s first block, belongs to $V$. This occurs because 1) if $j$ is unclustered when the variable $k$ reaches $I_k = I$, then $j$ will be clustered by Algorithm \ref{algo:homo}; 2) otherwise, $j$ is clustered by $c(j)$ which is placed at least two blocks prior.
        
        \item Otherwise, $I$ is selected via an else condition. Assume $\card{I} \geq 2$. The initial state of $I$ automatically makes $HC(I)$ true due to the interval definition. This implies that several clients in $I$ are already clustered by the time $k$ reaches $I_k = I$. Let the first block of $I$ at the initial state as $B$, with $b = I \setminus B$. For a state $I'$ just before $k$ reaches $I_k = I$, let $B'$ be the first block of $I'$ and $b' = I' \setminus B'$. Hence, any client $j \in b \setminus b'$ also satisfies $j \in V$. Given $HC(I')$ is false, the following holds:
        \begin{align}
            \sum_{j \in b \setminus b'} C_j &\geq \sum_{j \in b \setminus b'} (C^*_j - (\gamma-1)F^*_j) \notag \\
            &= C^*(b)-C^*(b') - (\gamma-1)(F^*(b)-F^*(b')) \notag \\
            &\geq C^*(b)-C^*(b') - K_4(F^*(b)-F^*(b')) \notag \\
            &\geq C^*(b)-K_4(F^*(B')+F^*(b')) - K_4(F^*(b)-F^*(b')) \notag \\
            &= C^*(b)-K_4(F^*(B')+F^*(b)) \notag \\
            &\geq C^*(b)-K_4(F^*(B)+F^*(b)) \notag \\
            &\geq \left( 1-\frac{K_4}{K_3} \right) C^*(b) = \left( 1-\frac{K_4}{K_3} \right) R(I). \notag
        \end{align}
        Note that it holds for even an interval of $\card{I} = 1$, since $R(I)=0$.
    \end{enumerate}

    Therefore, for any interval $I$,
    \[ \sum_{j \in V \cap I} C_j \geq \left( 1-\frac{K_4}{K_3} \right) R(I). \]
    Therefore, the total rerouting cost is at most
    \begin{align}
        \sum_{j \in \mathcal{C}} cost_{c(j)}(j) &\leq \sum_{j \in V} ((1-\eps')C_j + (3-\gamma)M_j + (\gamma-1)D_j) \notag \\
        &\quad+ \sum_{j \in \mathcal{C} \setminus V} (C_j + (3-\gamma)M_j + (\gamma-1)D_j) \notag \\
        &\leq \sum_{j \in \mathcal{C}} \left( \left( 1- \frac{\sum\limits_{I \in J} R(I)}{\sum\limits_{j' \in \mathcal{C}} C_{j'}} \cdot \left( 1-\frac{K_4}{K_3} \right) \eps' \right) C_j + (3-\gamma)M_j + (\gamma-1)D_j \right). \notag
    \end{align}
\end{proof}

In conclusion, it suffices to find a set of non-overlapping intervals $J$ which have a high $R(J)$ value. Here, we will briefly touch on the idea. We will iterate through the blocks in reverse order $(B_N, \ldots, B_0)$, cutting out suitable ranges that satisfy the interval conditions. Fix a point $r$ to be the right end of the interval, and expand the left end of the current range one block to the left until the range is suitable for processing. Suppose, at some point, the $C/F$ value of the current range is less than $K_2$, which is strictly less ratio of the input instance, $K_1$. Then, this range can be considered a minor part and can be excluded as there is no need for it to become an interval. Therefore, we only consider cases where the current range has a $C/F$ value of at least $K_2$.

However, in the situation where the current range's reward is small, meaning that the first block must avoid most of the $C^*$. In this case, if we keep expanding the range to the left, we will eventually reach the initial block $B_0$, satisfying the interval condition. There remains a subtle issue of the range's length exceeding $2L$ during the expansion process, but this implies that the $C^*$ value on the left side of the range is always exponentially increasing compared to the right side. This can be resolved by appropriately reducing the right side of the range.

\begin{lemma}
    For any support graph $G$ that is connection-dominant, $C^* > K_1F^*$, then the set of non-overlapping intervals $J$ obtained by Algorithm~\ref{algo:interval} satisfies $\sum_{I \in J} R(I) \geq \frac{1}{10^5} C^*$.
    \label{lemma:a3-perf}
\end{lemma}
\begin{algorithm}
\caption{\textsc{cutinterval}: Find a set of non-overlapping intervals $J$ with large rewards}
\begin{algorithmic}
\REQUIRE Network \(G = ((\mathcal{F}, \mathcal{C}), E)\).
\STATE $J := \emptyset$
\STATE $r := N$
\WHILE{$r > 0$}
    \STATE $l := r$
    \STATE $c := 0, f := 0$
    \WHILE{$l \geq 0$}
        \IF{$c \geq K_3 (f + F^*(B_l))$ and $C^*(B_l) \leq \frac{K_2-K_3}{K_2} c$}
            \STATE $J \gets J \cup \{[ l, r ]\}$
            \STATE $r \gets l-1$
            \STATE \textbf{break}
        \ENDIF
        \STATE $c \gets c + C^*(B_l), f \gets f + F^*(B_l)$
        \IF{$c < K_2 f$}
            \STATE $r \gets l-1$
            \STATE \textbf{break}
        \ENDIF
        \IF{$r-l+1 = 2L$}
            \STATE $c \gets c - \sum\limits_{i=l+L}^{r} C^*(B_i), f \gets f - \sum\limits_{i=l+L}^{r} F^*(B_i)$
            \STATE $r \gets r-L$
        \ENDIF
        \STATE $l \gets l-1$
    \ENDWHILE
\ENDWHILE
\STATE Include in $J$ every block (as an interval of size $1$) not already part of $J$
\RETURN $J$
\end{algorithmic}
\label{algo:interval}
\end{algorithm}

\begin{proof}
    The intervals of size $1$ are not in our interests. Thus in this proof, $J$ denotes the state before the last statement of Algorithm \ref{algo:interval} executed. First, $J$ is modified solely within the first if-statement, whose condition is a sufficient condition to be the interval, ensuring that $J$ comprises valid intervals. Consider every moment that variable $r$ is changed, with $r_i$ denoting its value prior to modification and $r_f$ post-modification. Let $P$ be the set of ranges $\left[ r_i+1, r_f \right]$ for all instances where $r$ is modified during the execution of Algorithm \ref{algo:interval}. The union of all elements in $P$ is exactly $\left[ 0, N \right]$, because (1) $l$ is decremented by $1$ consistently, and (2) when $l$ is reduced to $0$, the process transitions to the first if-statement, as $C^*(B_0) = 0$ and $F^*(B_0) = 0$. Also note that $J$ is a subset of $P$.

    For any range $\left[ x, y \right] \in P$ produced by the first or second if-statement, there could be multiple intervals from $P$ created by the third if-statement immediately preceding it. Let these be $\left[ y+1, y+L \right]$, $\left[ y+L+1, y+2L \right]$, $\ldots$, $\left[ y+(m-1)L+1, y+mL \right]$. To avoid entering first if-statement by neglecting the $c \geq K_3 (f + F^*(B_l))$ condition, for $1 \leq i \leq m-1$ and $(i-1)L+1 \leq k \leq iL$,
    \[ \sum_{j=y+k+1}^{y+(i+1)L} C^*(B_j) < K_3 \sum_{j=y+k}^{y+(i+1)L} F^*(B_j).  \]
    To avoid entering second if-statement, for $1 \leq i \leq m-1$ and $(i-1)L+1 \leq k \leq iL$,
    \[ \sum_{j=y+k}^{y+(i+1)L} C^*(B_j) \geq K_2 \sum_{j=y+k}^{y+(i+1)L} F^*(B_j). \]
    When both of the above two scenarios occur simultaneously,
    \[ \sum_{j=y+k+1}^{y+(i+1)L} C^*(B_j) < K_3 \sum_{j=y+k}^{y+(i+1)L} F^*(B_j) \leq \frac{K_3}{K_2} \sum_{j=y+k}^{y+(i+1)L} C^*(B_j) \]
    which implies
    \[ C^*(B_{y+k}) > \frac{K_2-K_3}{K_2} \cdot \sum_{j=y+k+1}^{y+(i+1)L} C^*(B_j) \geq \frac{K_2-K_3}{K_2} \cdot \sum_{j=y+iL+1}^{y+(i+1)L} C^*(B_j). \]
    The same result is achieved when another condition of the first if-statement is neglected.
    Hence, for $1 \leq i \leq m-1$,
    \[ \sum_{j=y+(i-1)L+1}^{iL} C^*(B_j) \geq \frac{(K_2-K_3)L}{K_2} \cdot \sum_{j=iL+1}^{(i+1)L} C^*(B_j). \]
    Note that if $m \geq 1$, then $y-x \geq L$ according to the logic of Algorithm \ref{algo:interval}, the following holds:
    \[ \sum_{j=x}^{y} C^*(B_j) \geq \frac{(K_2-K_3)L}{K_2} \cdot \sum_{j=y+1}^{y+L} C^*(B_j). \]
    Therefore, regardless of whether $m \geq 1$ or $m = 0$,
    \begin{align}
        \sum_{j=x}^{y+mL} C^*(B_j) &\leq \left( 1 + \frac{K_2}{(K_2-K_3)L} + \ldots + \left( \frac{K_2}{(K_2-K_3)L} \right)^m \right) \cdot \sum_{j=x}^{y} C^*(B_j) \notag \\
        &\leq \left( 1-\frac{K_2}{(K_2-K_3)L} \right)^{-1} \sum_{j=x}^{y} C^*(B_j). \notag
    \end{align}

    Consider the following two cases:
    \begin{enumerate}
        \item When $\left[ x, y \right] \in P$ is made by the first if-statement, the reward of $\left[ x, y \right] \in J$ is at least
        \begin{align}
            \sum_{j=x+1}^{y} C^*(B_j) &= -C^*(B_x) + \sum_{j=x}^{y} C^*(B_j) \notag \\
            &\geq -C^*(B_x) + \left( 1-\frac{K_2}{(K_2-K_3)L} \right) \cdot \sum_{j=x}^{y+mL} C^*(B_j) \notag \\
            &\geq \left( 1-\frac{K_2}{(K_2-K_3)L}-\frac{K_2-K_3}{K_2} \right) \cdot \sum_{j=x}^{y+mL} C^*(B_j). \notag
        \end{align}
        \item Otherwise, the total interval $\left[ x, y+mL \right]$ is facility-dominant. Precisely, the sum of $C^*(B_j)$ for $x \leq j \leq y+mL$ is bounded as follow:
        \begin{align}
            \sum_{j=x}^{y+mL} C^*(B_j) &\leq \left( 1-\frac{K_2}{(K_2-K_3)L} \right)^{-1} \sum_{j=x}^{y} C^*(B_j) \notag \\
            &< K_2 \left( 1-\frac{K_2}{(K_2-K_3)L} \right)^{-1} \sum_{j=x}^{y} F^*(B_j) \notag \\
            &\leq K_2 \left( 1-\frac{K_2}{(K_2-K_3)L} \right)^{-1} \sum_{j=x}^{y+mL} F^*(B_j) \notag \\
            &< K_1 \sum_{j=x}^{y+mL} F^*(B_j). \notag
        \end{align}
    \end{enumerate}

    Therefore, the ranges produced by the first if-statement and its preceding intervals from the third if-statement are only connection-dominant. Let the sum of the $C^*$ value of these ranges be $c^*$, and the sum of the $F^*$ values be $f^*$. From the assumption, $C^* > K_1F^*$. Also, by the above analysis, $(C^*-c^*) < K_2 \left( 1-\frac{K_2}{(K_2-K_3)L} \right)^{-1} (F^*-f^*)$. Then $c^*$ is at least
    \begin{align}
        c^* > C^* - K_2 \left( 1-\frac{K_2}{(K_2-K_3)L} \right)^{-1} (F^*-f^*) &\geq C^* - K_2 \left( 1-\frac{K_2}{(K_2-K_3)L} \right)^{-1} F^* \notag \\
        &> \left( 1- \frac{K_2}{K_1} \left( 1-\frac{K_2}{(K_2-K_3)L} \right)^{-1} \right) C^*. \notag
    \end{align}
    From the above analysis, $R(J)$ is at least
    \begin{align}
        \sum_{I \in J} R(I) &\geq \left( 1- \frac{K_2}{K_1} \left( 1-\frac{K_2}{(K_2-K_3)L} \right)^{-1} \right) c^* \notag \\
        &\geq \left( 1-\frac{K_2}{(K_2-K_3)L}-\frac{K_2-K_3}{K_2} \right) \cdot \left( 1- \frac{K_2}{K_1} \left( 1-\frac{K_2}{(K_2-K_3)L} \right)^{-1} \right) C^* \notag \\
        &\geq \frac{1}{10^5} C^* \notag
    \end{align}
\end{proof}

By directly applying Lemma \ref{lemma:a2-perf} and Lemma \ref{lemma:a3-perf}, Theorem \ref{thm:new-algo-perf} can be proved.

\begin{proof}
    By Lemma \ref{lemma:a2-perf} and Lemma \ref{lemma:a3-perf}, the rerouting cost of the clustering made by \textsc{conn}(\textsc{cutinterval}($G$)) (Algorithm~\ref{algo:conn} and~\ref{algo:interval}) is at most
    \begin{align}
        \sum_{j \in \mathcal{C}} cost_{c(j)}(j) &\leq \sum_{j \in \mathcal{C}} \left( \left( 1-\frac{C^*}{10^5 \sum\limits_{j' \in \mathcal{C}} C_{j'}} \eps_4 \right) C_j + (3-\gamma)M_j + (\gamma-1)D_j \right) \notag \\
        &\leq \sum_{j \in \mathcal{C}} \left( \left( 1-\frac{\eps_4}{10^5} \right) C_j + (3-\gamma)M_j + (\gamma-1)D_j \right). \notag
    \end{align}
\end{proof}

\section{Improved Bifactor Approximation}
\label{sec:proof-main}

In this section, we present an improved bifactor approximation algorithm for UFL, proving Theorem~\ref{thm:main-bifactor}. To deal with facility-dominant instances, we employ the JMS algorithm \cite{jain02}, which is known to be $(1.11, 1.7764)$-approximation algorithm.

\begin{algorithm}
\caption{Overall bi-factor clustering process}
\begin{algorithmic}
    \STATE Derive $(\Bar{x}, \Bar{y})$ from the optimal LP primal solution $(x^*, y^*)$.
    \IF{$C^* \leq K_1 F^*$}
        \STATE Execute the $(1.11, 1.7764)$-approximation algorithm.
    \ELSE
        \STATE Proceed with \textsc{conn}(\textsc{cutinterval}($G$)) (Algorithm~\ref{algo:conn} and~\ref{algo:interval}).
        \FORALL{cluster centers $j$}
            \STATE Open exactly one nearby facility with a probability of $\Bar{x}_{ij}$.
        \ENDFOR
        \FORALL{facilities $i$ not close to any cluster center}
            \STATE Open each facility with a probability of $\Bar{y}_i$ independently.
        \ENDFOR
        \STATE Connect each client to the nearest open facility.
    \ENDIF
\end{algorithmic}
\label{algo:final-bifactor}
\end{algorithm}

\begin{lemma}
    \cite{byrka10} Let $r \in \{0,1\}^{\card{\mathcal{F}}}$ be a vector indicating the opening status of each facility in Algorithm \ref{algo:final-bifactor}. For any subset $A \subseteq \mathcal{F}$ with $\sum_{i \in A} \Bar{y}_i > 0$, for a client $j \in \mathcal{C}$, the following holds:
    \[ E \left[ \min_{i \in A, r_i=1} d(i,j) \,\vert\, \sum_{i \in A} r_i \geq 1 \right] \leq d(j, A). \]
    \label{lemma:ub-closest-facility}
\end{lemma}

Now, we will show that Algorithm~\ref{algo:final-bifactor} satisfies Theorem~\ref{thm:main-bifactor}. Let $\gamma_0 \leq 1.6774$ be a solution of the below equation.
\[ \frac{1}{e}+\frac{1}{e^\gamma} - (\gamma-1)(1-\frac{1}{e} + (1-\eps_5)\frac{1}{e^\gamma}) = 0. \]

\bifactor*

\begin{proof}
    If $C^* \leq K_1 F^*$, then $1.11F^* + 1.7764C^* \leq 1.6774F^* + (1+2e^{-1.6774}-\eps_6)C^*$. Otherwise, a total cost is the cost of the output of \textsc{conn}(\textsc{cutinterval}($G$)). The expected cost of facility opening is given by
    \[ E[F_{SOL}] = \sum_{i \in \mathcal{F}} f_i \Bar{y}_i = \gamma \cdot \sum_{i \in \mathcal{F}} f_i y^*_i = \gamma \cdot F^*. \]
    
    Let $p_c$ be the probability that at least one close facility is opened, $p_d$ be the probability that at least one close/distant facility is opened, and $p_s$ be the probability that no adjacent facility is opened. As facilities are opened independently, it follows that $p_c \geq 1-e^{-1}$, $p_c+p_d \geq 1-e^{-\gamma}$. Thus, by Theorem \ref{thm:new-algo-perf} and Lemma \ref{lemma:ub-closest-facility}, the expected connection cost is at most
    \begin{align}
        \mathbb{E}[C_{SOL}] &\leq \sum_{j \in \mathcal{C}} (p_c \cdot C_j + p_d \cdot D_j + p_s \cdot ((1-\eps_5)C_j + (3-\gamma)M_j + (\gamma-1)D_j)) \notag \\
        &\leq \sum_{j \in \mathcal{C}} ((p_c + (1-\eps_5)p_s) \cdot C_j + (p_d+2p_s) \cdot D_j) \notag \\
        &= (1 + (2-\eps_5)p_s)C^* + \sum_{j \in \mathcal{C}} (F^*_j \cdot r_\gamma(j) \cdot (p_d+2p_s-(\gamma-1)(p_c+(1-\eps_5)p_s))) \notag \\
        &\leq \left (1 + \frac{2-\eps_5}{e^\gamma} \right) C^* + \sum_{j \in \mathcal{C}} F^*_j \cdot r_\gamma(j) \cdot \left( \frac{1}{e}+\frac{1}{e^\gamma} - (\gamma-1)(1-\frac{1}{e} + (1-\eps_5)\frac{1}{e^\gamma}) \right). \notag
    \end{align}
    Therefore, when $\gamma = 1.6774$, Algorithm \ref{algo:final-bifactor} is guaranteed to be $(1.6774, 1+2e^{-1.6774}-\eps_6)$-approximation algorithm.
\end{proof}

\section{Improved Unifactor Approximation}
\label{sec:uni}

In this section, we propose the algorithm that guarantees better unifactor approximation suggested by Li \cite{li13}, proving Theorem~\ref{thm:main-unifactor}.

\paragraph{Framework of \cite{li13}.}
Li showed that a hard instance for a certain $\gamma$ might not be a hard instance for another value of $\gamma$. This suggests that selecting a $\gamma$ value at random could improve the expected performance of the algorithm. 

They introduced a \textit{characteristic function} $h: (0, 1] \rightarrow \mathbb{R}$ to represent the distribution of distances between a client and its neighboring facilities. For a client $j \in \mathcal{C}$, assume $i_1, i_2, \ldots, i_k$ are the facilities within $\mathcal{C}_j \cup \mathcal{D}_j$, ordered by increasing distance from $j$. Then, for $0 < p \leq 1$, $h_j(p)$ is defined as $d(j, i_t)$, where $t$ is the smallest index satisfying $\sum_{l=1}^{t} y^*_{i_l} \geq p$. Also, they improved the bound of the rerouting cost as $d(j, \mathcal{C}_{j'} \setminus (\mathcal{C}_j \cup \mathcal{D}_j)) \leq (2-\gamma)M_j + (\gamma-1)D_j + C_{j'} + M_{j'}$. 
For fixed $\gamma$, the expected connection cost for $j$ for the aforementioned algorithm of~\cite{byrka10} is at most 
    \[ \mathbb{E}[C_j] \leq \int_{0}^{1} h_j(p) e^{-\gamma p} \gamma \,dp + e^{-\gamma} \left( \gamma \int_{0}^{1} h_j(p) \,dp + (3-\gamma) h_j(\frac{1}{\gamma}) \right). \]

Therefore, it can be modeled as a $0$-sum game to analyze the approximation ratio. 
The characteristic function for the whole instance is given by $h(p) = \sum_{j \in \mathcal{C}} h_j(p)$. 
Assuming $h$ is normalized so that $\int_{0}^1 h(p) \,dp = 1$, the algorithm proceeds as follows: with probability $\kappa$, it employs the JMS algorithm \cite{jain02}. Mahdian \cite{mahdian02} proved that the JMS algorithm achieves a $(1.11, 1.7764)$-approximation. Otherwise, $\gamma$ is sampled randomly from the distribution $\mu: (1, \infty] \rightarrow \mathbb{R}^*$, ensuring that $\kappa + \int_{1}^{\infty} \mu(\gamma) \, d\gamma = 1$. Thus, the value of the $0$-sum game, i.e., the approximation ratio of the algorithm under a fixed strategy $(\kappa, \mu)$, is calculated as follows:
\[ \nu (\kappa, \mu, h) = \max \Bigl\{ \int_{1}^{\infty} \gamma \mu(\gamma) \,d\gamma + 1.11\kappa, \int_{1}^{\infty} \alpha(\gamma, h) \mu(\gamma) \,d\gamma + 1.7764\kappa \Bigr\} \]
where
\[ \alpha(\gamma, h) = \int_{0}^{1} h(p) e^{-\gamma p} \gamma \,dp + e^{-\gamma} \left( \gamma + (3-\gamma) h\left( \frac{1}{\gamma} \right) \right). \]

Moreover, for a given probability density function $\mu$ for $\gamma$, it can be shown that the characteristic function for the hardest instance is a \textit{threshold function}, which defined as
\[ h_q(p) =
\begin{cases}
    \frac{1}{1-q}, &\text{for } p > q \\
    0, &\text{for } p \leq q \\
\end{cases} \]
for some $0 \leq q < 1$. This means that the final approximation ratio for some $\mu$ is given by $\max_{0 \leq q < 1} \nu (\kappa, \mu, h_q)$. In \cite{li13}, the suggested distribution for $\alphali$ is $\mu(p) = \theta D(p-\gamma_1) + \frac{1-\kappa-\theta}{\gamma_2-\gamma_1} [\gamma_1 < p < \gamma_2]$, where $D$ is Dirac delta function, $\gamma_1 = 1.479311$, $\gamma_2 = 2.016569$, $\theta = 0.503357$, $\kappa = 0.195583$.

\paragraph{Our Improvement.}

Let $h: (0, 1] \rightarrow \mathbb{R}^*$ be the characteristic function of a given instance. Using Algorithm \ref{algo:final-bifactor}, the expected connection cost for some $\gamma$ is given as follows:

\begin{lemma}
     If $1.6 < \gamma < 2$, then for any connection-dominant instance such that $C^* > K_1F^*$ and $\eps_7 = 2 \times 10^{-42}$, the expected connection cost is at most
    \[ \mathbb{E}[C] \leq \int_{0}^{1} h(p) e^{-\gamma p} \gamma \,dp + e^{-\gamma} \left( \gamma (1-\eps_7) \int_{0}^{1} h(p) \,dp + (3-\gamma) h(\frac{1}{\gamma}) \right). \]
    \label{lemma:char-func}
\end{lemma}
\begin{proof}
    For any client $j$, if there exists at least one facility opened in $\mathcal{C}_j \cup \mathcal{D}_j$, the connection cost for $j$ is given by
    \[ \int_{0}^1 h_j(p) e^{-\gamma p} \gamma \,dp, \]
    in the same way showed by Li \cite{li13}.
    
    For the connection-dominant instance, our algorithm gives an improved rerouting cost. Therefore, by Theorem \ref{thm:new-algo-perf}, the expected connection cost is at most
    \begin{align}
        \mathbb{E}[C] &\leq \int_{0}^{1} h(p) e^{-\gamma p} \gamma \,dp \notag \\
        &\quad+ e^{-\gamma} \left( \gamma (1-\eps_5) \int_{0}^{1/\gamma} h(p) \,dp + \frac{\gamma}{\gamma-1} \int_{1/\gamma}^{1} h(p) \,dp + (3-\gamma) h(\frac{1}{\gamma}) \right). \notag
    \end{align}
    Also, from the $C^* > K_1F^*$ condition, the following holds:
    \[ \frac{\sum_{j \in \mathcal{C}} C_j}{\sum_{j \in \mathcal{C}} D_j} = \frac{\sum_{j \in \mathcal{C}} (C^*_j - (\gamma-1)r_\gamma(j) F^*_j)}{\sum_{j \in \mathcal{C}} (C^*_j + r_\gamma(j)F^*_j)} \geq \frac{\sum_{j \in \mathcal{C}} (C^*_j - (\gamma-1) F^*_j)}{\sum_{j \in \mathcal{C}} (C^*_j + F^*_j)} \geq \frac{K_1-\gamma+1}{K_1+1} \]
    which implies
    \[ \frac{\gamma \int_{0}^{1/\gamma} h(p) \,dp}{\frac{\gamma}{\gamma-1} \int_{1/\gamma}^{1} h(p) \,dp} = \frac{\sum_{j \in \mathcal{C}} C_j}{\sum_{j \in \mathcal{C}} D_j} \geq \frac{K_1-\gamma+1}{K_1+1}. \]

    Therefore,
    \[ \mathbb{E}[C] \leq \int_{0}^{1} h(p) e^{-\gamma p} \gamma \,dp + e^{-\gamma} \left( \gamma (1 - \frac{K_1-\gamma+1}{2K_1-\gamma+2} \cdot \eps_5) \int_{0}^{1} h(p) \,dp + (3-\gamma) h(\frac{1}{\gamma}) \right). \]
\end{proof}

From the above lemma, we define a new $0$-sum game value as follows, assuming $h$ is scaled up such that $\int_{0}^1 h(p) \,dp = 1$.
\[ \nu' (\mu, \theta, h) = \max \Bigl\{ \int_{1}^{\infty} \gamma \mu(\gamma) \,d\gamma + 1.11\kappa, \int_{1}^{\infty} \alpha'(\gamma, h) \mu(\gamma) \,d\gamma + 1.7764\kappa \Bigr\} \]
where
\[
    \alpha'(\gamma, h) = 
    \begin{cases}
        \int_{0}^{1} h(p) e^{-\gamma p} \gamma \,dp + e^{-\gamma} \left( \gamma (1-\eps_7) + (3-\gamma) h(\frac{1}{\gamma}) \right), & \text{if $1.6 < \gamma <2$} \\
        \int_{0}^{1} h(p) e^{-\gamma p} \gamma \,dp + e^{-\gamma} \left( \gamma + (3-\gamma) h(\frac{1}{\gamma}) \right). & \text{otherwise} \\
    \end{cases}
\]

Note that $\alpha'$ is still linear for $h$. Therefore, even if the game definition changes, the adversary's choice of the characteristic function $h$ remains a threshold function $h_q$ for some $0 \leq q < 1$. Given that there is a positive probability of sampling $\gamma$ between $1.6$ and $2$, it is possible to achieve a lower cost.

\begin{lemma}
    Let $\mu_2(\gamma) = (1-\eps_7)\mu_1(\gamma) + \eps_7(1-\kappa_2) D(\gamma-1)$, where $D$ is a Dirac-delta function. Then the following holds:
    \begin{align}
        &\max_{0 \leq q < 1} \max \Bigl\{ \int_{1}^{\infty} \gamma \mu_2(\gamma) \,d\gamma + 1.11\kappa_2, \int_{1}^{\infty} \alpha'(\gamma, h_q) \mu_2(\gamma) \,d\gamma + 1.7764\kappa_2 \Bigr\} \notag \\
        < &\max_{0 \leq q < 1} \max \Bigl\{ \int_{1}^{\infty} \gamma \mu_1(\gamma) \,d\gamma + 1.11\kappa_2, \int_{1}^{\infty} \alpha(\gamma, h_q) \mu_1(\gamma) \,d\gamma + 1.7764\kappa_2 \Bigr\} - \frac{\eps_7}{1000}. \notag
    \end{align}
    \label{lemma:mu2-perf}
\end{lemma}
\begin{proof}
    For any $0 \leq q < 1$,
    \[ \int_{1}^{\infty} h_q(p) e^{-\gamma p} \gamma \,dp = \frac{e^{-qx}-e^{-x}}{1-q}. \]

    Since $\int_{1.6}^{2} \mu_1(\gamma) \,d\gamma > 0.01$,
    \begin{align}
        \int_{1}^{\infty} \alpha'(\gamma, h_q) \mu_2(\gamma) \,d\gamma &= \int_{1}^{\infty} \alpha(\gamma, h_q) \mu_2(\gamma) \,d\gamma - \eps_7 \int_{1.6}^{2} \gamma e^{-\gamma} \mu_2(\gamma) \,d\gamma \notag \\
        &< \int_{1}^{\infty} \alpha(\gamma, h_q) \mu_2(\gamma) \,d\gamma - \frac{\eps_7}{100} \cdot \frac{1.6}{e^{1.6}}. \notag
    \end{align}

    Also, the following holds:
    \begin{align}
        \int_{1}^{\infty} \gamma \mu_2(\gamma) \,d\gamma &= (1-\eps_7) \int_{1}^{\infty} \gamma \mu_1(\gamma) \,d\gamma + \eps_7 (1-\kappa_2) \notag \\
        &< \int_{1}^{\infty} \gamma \mu_1(\gamma) \,d\gamma - \eps_7 \cdot (1.487-1.11\kappa_2) + \eps_7 (1-\kappa_2) \notag \\
        &< \int_{1}^{\infty} \gamma \mu_1(\gamma) \,d\gamma - 0.377\eps_7. \notag
    \end{align}

    Therefore,
    \begin{align}
        &\max \Bigl\{ \int_{1}^{\infty} \gamma \mu_2(\gamma) \,d\gamma + 1.11\kappa_2, \int_{1}^{\infty} \alpha'(\gamma, h_q) \mu_2(\gamma) \,d\gamma + 1.7764\kappa_2 \Bigr\} \notag \\
        < &\max \Bigl\{ \int_{1}^{\infty} \gamma \mu_1(\gamma) \,d\gamma + 1.11\kappa_2, \int_{1}^{\infty} \alpha(\gamma, h_q) \mu_1(\gamma) \,d\gamma + 1.7764\kappa_2 \Bigr\} - \frac{\eps_7}{1000} \notag \\
        \leq \max_{0 \leq q < 1} &\max \Bigl\{ \int_{1}^{\infty} \gamma \mu_1(\gamma) \,d\gamma + 1.11\kappa_2, \int_{1}^{\infty} \alpha(\gamma, h_q) \mu_1(\gamma) \,d\gamma + 1.7764\kappa_2 \Bigr\} - \frac{\eps_7}{1000} \notag
    \end{align}
    which implies that the maximum value of the new $0$-sum game has strictly less value than the original one.
\end{proof}

Therefore, we present an improved unifactor approximation algorithm, Algorithm~\ref{algo:final-uni}. 
\begin{algorithm}
\caption{Overall uni-factor clustering process}
\begin{algorithmic}
    \STATE Derive $(\Bar{x}, \Bar{y})$ from the optimal LP primal solution $(x^*, y^*)$.
    \STATE Let $X$ be a random variable that takes the value 1 with probability $1-\kappa_2$, and 0 otherwise.
    \IF{$C^* \leq K_1 F^*$}
        \STATE Execute the $(1.11, 1.7764)$-approximation algorithm.
    \ELSIF{X = 1}
        \STATE Execute the $(1.11, 1.7764)$-approximation algorithm.
    \ELSE
        \STATE Takes $\gamma$ from the distribution $\mu_2(\gamma) = (1-\eps_7)\mu_1(\gamma) + \eps_7(1-\kappa_2) D(\gamma-1)$.
        \IF{$1.6 \leq \gamma \leq 2$}
            \STATE Proceed with \textsc{conn}(\textsc{cutinterval}($G$)) (Algorithm~\ref{algo:conn} and~\ref{algo:interval}).
        \ELSE
            \STATE Proceed with \textsc{greedy}($G$).
        \ENDIF
        \FORALL{cluster centers $j$}
            \STATE Open exactly one nearby facility with a probability of $\Bar{x}_{ij}$.
        \ENDFOR
        \FORALL{facilities $i$ not close to any cluster center}
            \STATE Open each facility with a probability of $\Bar{y}_i$ independently.
        \ENDFOR
        \STATE Connect each client to the nearest open facility.
    \ENDIF
\end{algorithmic}
\label{algo:final-uni}
\end{algorithm}

When $C^* \leq K_1 F^*$, an inequality $1.11F^* + 1.7764C^* \leq 1.487F^* + 1.487C^*$ is satisfied. For the case $C^* > K_1 F^*$, by Lemma \ref{lemma:mu2-perf}, Algorithm \ref{algo:final-uni} shows an improved unifactor approximation with $\eps_8 = \frac{\eps_7}{1000} = 2 \times 10^{-45}$.

\section{APX-Hardness}
\label{sec:hardness}

In this section, we prove that UFL is APX-Hard in Euclidean spaces. We use the following result from Austrin, Khot, and Safra~\cite{austrin2011inapproximability}. For $\rho \in [-1, +1]$ and $\mu \in [0, 1]$, let $\Gamma_{\rho}(\mu) := \Pr[X \leq \Phi^{-1}(\mu) \wedge Y \leq \Phi^{-1}(\mu)]$ where $X,Y$ are standard Gaussian random variables with covariance $\rho$ and $\Phi$ is the cumulative density function of the standard normal distribution. 
\begin{theorem}
Assuming the Unique Games Conjecture, 
for any $q \in (0, 1/2)$ and $\eps > 0$, it is NP-hard to, given a graph $G = (V, E)$, distinguish between the following two cases.
\begin{itemize}
    \item (Completeness) $G$ contains an independent set of size $q |V|$. 
    \item (Soundness) For any $T \subseteq V$, the number of edges with both endpoints in $T$ is at least $|E| \cdot (\Gamma_{-q/(1-q)}(\mu)-\eps)$ where $\mu = |T| / |V|$. 
\end{itemize}
\end{theorem}

Fix an arbitrary $q \in (0, 1/2)$. 
Without loss of generality, assume $V = [n]$. Also let $m := |E|$. Our UFL instance has $V$ as the set of facilities and $E$ as the set of clients. The ambient Euclidean space is $\R^n$, and let $e_i$ be the $i$th standard unit vector (i.e., $(e_i)_i = 1$ and $(e_i)_j = 0$ for every $j \neq i$). Then each facility $i \in V$ is located at $e_i$ and each client $(i, j) \in E$ is located at $e_i + e_j$. Finally, let $\lambda$ be the common facility cost for every $i \in \calF$ to be determined. This finishes the description of the UFL instance.

In the completeness case, there is an independent set $U$ of size $qn$. We open $V \setminus U$. Since $V \setminus U$ is a vertex cover, every client in $E$ has a facility at distance $1$, so the total cost is 
\[
\lambda(1-q)n + m. 
\]
In the soundness case, consider any solution that opens $S \subseteq V$ and let $T = V \setminus S$ and $\mu = |T|/n$. By the soundness guarantee, at least $m(\Gamma_{-q/(1-q)} - \eps)$ clients do not have a facility at distance $1$. Since every client-facility distance is either $1$ or $\sqrt{3}$, the total cost is at least 
\begin{equation}
\lambda(1-\mu)n + 
m\big( 1 + (\sqrt{3} - 1)(\Gamma_{-q/(1-q)}(\mu) - \eps) 
\big).
\label{eq:soundness}
\end{equation}
For fixed $q$, the function $\Gamma_{-q/(1-q)}(\mu)$ is a strictly convex function of $\mu$, so if we let $\lambda$ such that
\[
- \lambda n + m(\sqrt{3} - 1) \frac{d\Gamma_{-q/(1-q)}(\mu)}{d\mu}|_{\mu=q} = 0, 
\]
then~\eqref{eq:soundness} is minimized when when $\mu = q$, which becomes 
\[
\lambda(1-q)n + 
m\big( 1 + (\sqrt{3} - 1)(\Gamma_{-q/(1-q)}(q) - \eps) 
\big).
\]
Furthermore, we can notice that 
\[
\lambda = \frac{m}{n} \cdot (\sqrt{3} - 1)\frac{d\Gamma_{-q/(1-q)}(\mu)}{d\mu}|_{\mu=q} = \Theta(\frac{m}{n}).
\]
Then one can see that the optimal value in the soundness case is at least 
$m(\sqrt{3} - 1)(\Gamma_{-q/(1-q)}(q) - \eps)$ larger than the optimal value in the completeness case. For a fixed $q$, by choosing $\eps$ sufficiently small, one can ensure that this excess is at least a $\delta$ fraction of the completeness case optimal value for some constant $\delta > 0$, which proves a $(1 + \delta)$-hardness of approximation.

\section{Conclusion}
The most natural open problem is to get an improved approximation for bifactor or unifactor approximation for UFL. Though we show a strict separation between general and Euclidean metrics for bifactor approximation in a certain regime, it is not achieved for all regimes of bifactor or unifactor approximation. 

Whereas our algorithm is based on the {\em primal rounding} approach of~\cite{byrka10} and~\cite{li13}, it might be a fruitful research direction to design a variant of the Jain-Mahdian-Saberi algorithm~\cite{jain02} (greedy algorithm analyzed by the dual fitting method) or Jain-Vazirani~\cite{jain2001approximation} (primal-dual algorithm) for a further improvement. In particular, as the best unifactor approximation for UFL in both general and Euclidean metrics employ the $(1.11, 1.7764)$-approximation of 
the JMS algorithm as a black box, improving the JMS algorithm will directly yield a better result for the best unifactor approximation for UFL. 
The JV algorithm was already improved in Euclidean spaces~\cite{ahmadian2019better, grandoni2022refined, cohen2022improved}, but they are not enough for UFL. 

\bibliography{references}

\newcommand{\etalchar}[1]{$^{#1}$}
\begin{thebibliography}{CAVGLS23}

\bibitem[AKS11]{austrin2011inapproximability}
Per Austrin, Subhash Khot, and Muli Safra.
\newblock Inapproximability of vertex cover and independent set in bounded
  degree graphs.
\newblock {\em Theory of Computing}, 7(1):27--43, 2011.

\bibitem[ANFSW19]{ahmadian2019better}
Sara Ahmadian, Ashkan Norouzi-Fard, Ola Svensson, and Justin Ward.
\newblock Better guarantees for k-means and euclidean k-median by primal-dual
  algorithms.
\newblock {\em SIAM Journal on Computing}, 49(4):FOCS17--97, 2019.

\bibitem[BA10]{byrka10}
Jaroslaw Byrka and Karen Aardal.
\newblock An optimal bifactor approximation algorithm for the metric
  uncapacitated facility location problem.
\newblock {\em SIAM Journal on Computing}, 39(6):2212--2231, 2010.

\bibitem[CAEMN22]{cohen2022improved}
Vincent Cohen-Addad, Hossein Esfandiari, Vahab Mirrokni, and Shyam Narayanan.
\newblock Improved approximations for euclidean k-means and k-median, via
  nested quasi-independent sets.
\newblock In {\em Proceedings of the 54th Annual ACM SIGACT Symposium on Theory
  of Computing}, pages 1621--1628, 2022.

\bibitem[CAFS21]{cohen2021near}
Vincent Cohen-Addad, Andreas~Emil Feldmann, and David Saulpic.
\newblock Near-linear time approximation schemes for clustering in doubling
  metrics.
\newblock {\em Journal of the ACM (JACM)}, 68(6):1--34, 2021.

\bibitem[CAK19]{cohen2019inapproximability}
Vincent Cohen-Addad and CS~Karthik.
\newblock Inapproximability of clustering in lp metrics.
\newblock In {\em 2019 IEEE 60th Annual Symposium on Foundations of Computer
  Science (FOCS)}, pages 519--539. IEEE, 2019.

\bibitem[CASL22]{cohen2022johnson}
Vincent Cohen-Addad, Karthik~C S, and Euiwoong Lee.
\newblock Johnson coverage hypothesis: Inapproximability of k-means and
  k-median in $\ell_p$-metrics.
\newblock In {\em Proceedings of the 2022 Annual ACM-SIAM Symposium on Discrete
  Algorithms (SODA)}, pages 1493--1530. SIAM, 2022.

\bibitem[CAVGLS23]{cohen2023breaching}
Vincent Cohen-Addad~Viallat, Fabrizio Grandoni, Euiwoong Lee, and Chris
  Schwiegelshohn.
\newblock Breaching the 2 lmp approximation barrier for facility location with
  applications to k-median.
\newblock In {\em Proceedings of the 2023 Annual ACM-SIAM Symposium on Discrete
  Algorithms (SODA)}, pages 940--986. SIAM, 2023.

\bibitem[GK99]{guha98}
Sudipto Guha and Samir Khuller.
\newblock Greedy strikes back: Improved facility location algorithms.
\newblock {\em Journal of algorithms}, 31(1):228--248, 1999.

\bibitem[GOR{\etalchar{+}}22]{grandoni2022refined}
Fabrizio Grandoni, Rafail Ostrovsky, Yuval Rabani, Leonard~J Schulman, and
  Rakesh Venkat.
\newblock A refined approximation for euclidean k-means.
\newblock {\em Information Processing Letters}, 176:106251, 2022.

\bibitem[GPST23]{gowda2023improved}
Kishen~N Gowda, Thomas Pensyl, Aravind Srinivasan, and Khoa Trinh.
\newblock Improved bi-point rounding algorithms and a golden barrier for
  k-median.
\newblock In {\em Proceedings of the 2023 Annual ACM-SIAM Symposium on Discrete
  Algorithms (SODA)}, pages 987--1011. SIAM, 2023.

\bibitem[JMS02]{jain02}
Kamal Jain, Mohammad Mahdian, and Amin Saberi.
\newblock A new greedy approach for facility location problems.
\newblock In {\em Proceedings of the thiry-fourth annual ACM symposium on
  Theory of computing}, pages 731--740, 2002.

\bibitem[JV01]{jain2001approximation}
Kamal Jain and Vijay~V Vazirani.
\newblock Approximation algorithms for metric facility location and k-median
  problems using the primal-dual schema and lagrangian relaxation.
\newblock {\em Journal of the ACM (JACM)}, 48(2):274--296, 2001.

\bibitem[Li13]{li13}
Shi Li.
\newblock A 1.488 approximation algorithm for the uncapacitated facility
  location problem.
\newblock {\em Information and Computation}, 222:45--58, 2013.

\bibitem[MYZ02]{mahdian02}
Mohammad Mahdian, Yinyu Ye, and Jiawei Zhang.
\newblock A 1.52-approximation algorithm for the uncapacitated facility
  location problem.
\newblock In {\em Proc. of APPROX}, pages 229--242, 2002.

\bibitem[Ran55]{rankin55}
Robert~Alexander Rankin.
\newblock The closest packing of spherical caps in n dimensions.
\newblock {\em Glasgow Mathematical Journal}, 2(3):139--144, 1955.

\end{thebibliography}

\appendix

\section{Appendix}

\subsection{Postponed Proof of Lemma \ref{lemma:big-remote-arm}}
At this point, the lemma is equivalent to show that the above inequalities can not still holds simultaneously when $x = K_6$. Recall that $1.6 < \gamma < 2$, $\theta \leq k \leq \frac{1+\delta}{2}$, $l \leq 1-k-\delta$, $0 \leq r \leq 1$. Also, by Lemma \ref{lemma:lb-reroute-vector}, $0.99k < l$ holds too.

Denote the uncertainty of variable (or function) $X$ as $\Delta X$. It means that for given value $X_0$, variable $X$ may have $[X_0-\Delta X, X_0 + \Delta X]$. Then the following holds, which enables the composition between variables:
\[ \Delta(A+B) = (A+\Delta A + B + \Delta B) - (A+B) = \Delta A + \Delta B, \quad \Delta(tA) = \vert t \vert \Delta A, \]
\[ \Delta(AB) = (A+\Delta A)(B+\Delta B) - AB = A \Delta B + B \Delta A + \Delta A \Delta B. \]

Let $W = \gamma-1+l-k+\delta$. The first one can be rewritten as $(A_1 K_6+B_1)^2 + (A_2 K_6+B_2)^2 - \frac{W^2 (K_6+1)^2}{0.995} < 0$, where
\[ A_1 = (\gamma-1)(3-2l) + (l-k+\delta)(2\gamma-3)-\eps_1 W, \quad B_1 = (\gamma-1)r \cdot ((1+\eps_1)W + (2-k+\delta)(2-\gamma)), \]
\[ A_2 = 2(\gamma-1)l, \quad B_2 = (\gamma-1)(2-\gamma)rl. \]

\begin{enumerate}
    \item Uncertainty of $A1$.
    \[ \Delta((\gamma-1)(3-2l)) \leq \Delta(3-2l) + 3\Delta(\gamma-1) + 2d^2 \leq 5d+2d^2, \]
    \[ \Delta((l-k+\delta)(2\gamma-3)) \leq \Delta(2\gamma-3) + \Delta(l-k) + 4d^2 \leq 4d+4d^2, \]
    \[ \Delta(W) \leq 3d. \]
    Therefore,
    \[ \Delta(A_1) \leq (9+3\eps_1)d + 6d^2 < 9.1d. \]

    \item Uncertainty of $B1$.
    \[ \Delta ((\gamma-1)r) \leq \Delta(r) + \Delta(\gamma-1) + d^2 = 2d+d^2, \]
    \[ \Delta((2-k+\delta)(2-\gamma)) \leq 2\Delta(2-\gamma) + 0.4\Delta(2-k+\delta) + d^2 = 2.4d+d^2. \]
    Note that $0.59 < W < 2$. Therefore,
    \[ \Delta(B_1) \leq \Delta((1+\eps_1)W + (2-k+\delta)(2-\gamma)) + (2.8+2\eps_1) \Delta ((\gamma-1)r) + (2d+d^2)(2.4d+d^2) < 8.1d. \]
    
    \item Uncertainty of $A2$.
    \[ \Delta(A_2) = 2\Delta ((\gamma-1)l) \leq 2(\Delta(l) + \Delta(\gamma-1) + d^2) = 4d+2d^2 < 4.1d. \]

    \item Uncertainty of $B2$.
    \[ \Delta((\gamma-1)(2-\gamma)) \leq \Delta(2-\gamma) + 0.4\Delta(\gamma-1) + d^2 \leq 1.4d+d^2, \]
    \[ \Delta(rl) \leq 2d + d^2. \]
    Therefore,
    \[ \Delta(B_2) = \Delta((\gamma-1)(2-\gamma)rl) \leq 0.24 \cdot \Delta(rl) + \Delta((\gamma-1)(2-\gamma)) + (1.4d+d^2)(2d+d^2) \leq 1.9d. \]
\end{enumerate}

The range for coefficients are given as follow:
\[ A_1 < (3-2l) + (l-k+\delta) \leq 3, \quad B_1 \leq 1 \cdot 2(1+\eps_1) + 0.24 \cdot 2) < 2.5, \]
\[ A_2 \leq 2, \quad B_2 < 0.24 \cdot 1 = 0.24. \]
Thus, the uncertainty of the first inequality is at most
\begin{align}
    &\quad \,\, 2(A_1 K_6 + B_1) (K_6\Delta A_1 + \Delta B_1) \notag \\
    &\quad+ 2(A_2 K_6 + B_2) (K_6\Delta A_2 + \Delta B_2) + \frac{(K_6+1)^2}{0.995} 2W\Delta W + O(d^2) \notag \\
    &\leq 2(3 \cdot 1.302 + 2.5)(1.302 \cdot 9.1d + 8.1d) \notag \\
    &\quad+ 2(2 \cdot 1.302 + 0.24)(1.302 \cdot 4.1d + 1.9d) + \frac{12(K_6+1)^2}{0.995} d + O(d^2) \notag \\
    &\leq 360.8d. \notag
\end{align}

Similarly, the second inequality can be rewritten as $(A_3 K_6 + B_3)^2 + (A_4 K_6 + B_4)^2 - \frac{(K_6+1)^2}{0.995} < 0$, where
\begin{align}
    &A_3 = \frac{3-3\delta + 2\delta \gamma - 2l}{1+\delta}, \,\, B_3 = -(\gamma-1)r \cdot \left( \frac{-(3-\gamma)(1-\delta) + 2l}{1+\delta} + \eps_1 \right), \notag \\
    &\,\, A_4 = \frac{2l}{1+\delta}, \,\, B_4 = -\frac{2l(\gamma-1)r}{1+\delta}. \notag
\end{align}

\begin{enumerate}
    \item Uncertainty of $A3$ is given by $\Delta(A_3) = 2d$.

    \item Uncertainty of $B3$.
    \[ \Delta ((\gamma-1)r) \leq \Delta(r) + \Delta(\gamma-1) + d^2 = 2d+d^2, \]
    \[ \Delta(\frac{-(3-\gamma)(1-\delta) + 2l}{1+\delta}) < 3d. \]
    Therefore,
    \[ \Delta(B_3) < \Delta(\frac{-(3-\gamma)(1-\delta) + 2l}{1+\delta}) + (3-1.6) \Delta ((\gamma-1)r) + (2d+d^2) \cdot 3d < 5.9d. \]
    
    \item Uncertainty of $A4$ is given by $\Delta(A_4) < 2d$.

    \item Uncertainty of $B4$.
    \[ \Delta(rl) \leq 2d + d^2. \]
    Therefore,
    \[ \Delta(B_4) < 2(\Delta(\gamma-1) + \Delta(rl) + (2d+d^2) \cdot d) \leq 6.1d. \]
\end{enumerate}

The range for coefficients are given as follow:
\[ A_3 < 3, \quad B_3 \leq 1 \cdot (3-1.4) = 1.6, \quad A_4 \leq 2, \quad B_4 < 0. \]
Thus, the uncertainty of the second inequality is at most
\begin{align}
    &\quad \,\, 2(A_3 K_6 + B_3) (K_6\Delta A_3 + \Delta B_3) + 2(A_4 K_6 + B_4) (K_6\Delta A_4 + \Delta B_4) + O(d^2) \notag \\
    &\leq 2(3 \cdot 1.302 + 1.6)(1.302 \cdot 2d + 5.9d) + 2(2 \cdot 1.302 + 0)(1.302 \cdot 2d + 6.1d) + O(d^2) \notag \\
    &\leq 139.1d. \notag
\end{align}

\subsection{Parameter Setting}

This section enumerates the values of parameters discussed in the text, along with the sufficient conditions required to satisfy the theorems and lemmas presented. The parameters specified throughout this text are chosen to fulfill the following conditions:

\begin{itemize}
    \item $K_1 = 1.3025 > K_2 = 1.3024 > K_3 = 1.3023 > K_4 = 1.3022 > K_5 = 1.3021 > K_6 = 1.3020$.
    \item $1.6 < \gamma < 2$, $\theta = \frac{K_6+1-\gamma}{2K_6+2-\gamma}$, $\alpha = 5 \times 10^{-4}$, $M = 5 \times 10^6$, $r = 10^{-8}$, $L = 2 \times 10^8$.
    \item Note that for $1.6 < \gamma < 2$, $0.2336 \leq \theta \leq 0.3613$.
    \item $\eps_1 = 10^{-12}$, $\eps_2 = 5 \times 10^{-18}$, $\eps_3 = 3 \times 10^{-32}$, $\eps_4 = 2 \times 10^{-36}$, $\eps_5 = 2 \times 10^{-41}$, $\eps_6 = 3 \times 10^{-42}$, $\eps_7 = 2 \times 10^{-42}$, $\eps_8 = 2 \times 10^{-45}$, $\delta = 3 \times 10^{-23}$, $\delta' = 7 \times 10^{-32}$.
\end{itemize}
\begin{enumerate}
    \item From Lemma \ref{lemma:big-remote-arm},
    \[ (1-0.9995) \cdot 0.2319 \geq 2 \cdot 0.998 \cdot \sin{\alpha}. \]
    \item From Lemma \ref{lemma:lb-reroute-vector},
    \[ \frac{\eps_1 + \delta}{\theta} \leq \frac{1}{100}. \]
    \item From Lemma \ref{lemma:lb-cone-prob},
    \[ \frac{1}{0.98\theta r} \cdot \frac{\delta + \frac{\eps_1}{2}}{1+\delta} < \frac{0.00099\theta}{1-\theta}. \]
    \item From the definition of $\phi_r$, the bottleneck of $\phi_r$ is given at $x=2$:
    \[ 1^2 + 2^2 + 2 \cdot 1 \cdot 2 \cdot \cos{\phi_r} \leq (3-2r)^2. \]
    Thus, from Theorem \ref{thm:max-small-remote-arm},
    \[ \alpha - \phi_r > 0, \quad 2\phi_r < \frac{1}{100}, \]
    \[ 2(1+\delta) \cdot \sin{\alpha} \leq 0.98\theta, \quad M \geq \left( 1 + \frac{1}{\sin{(\alpha-\phi_r)}} \right) \cdot \frac{\log{\frac{1-\theta}{0.99\theta}}}{\log{1.001}}. \]
    \item From Theorem \ref{lemma:saving-expansion},
    \[ \frac{36}{25} (1+\delta) \leq \frac{3-\eps_1}{2(1+\delta)}, \quad 72\delta + 25\eps_1 \leq 1. \]
    \item From Theorem \ref{thm:geom-final},
    \[ (\delta+\eps_2) \left(\frac{2K_5+2-\gamma}{K_5-K_6} \cdot \frac{K_5}{K_5-\gamma+1}\right) \cdot \max \left( \frac{125M(1+\delta)}{2}, \frac{1}{\eps_1-\eps_2} \right) \leq 1. \]
    \item From Theorem \ref{thm:homogeneous-clustering},
    \[ \frac{(\delta + \eps_3)(2-\gamma+2K_4)}{K_4} \leq (\eps_2-\eps_3) \frac{(K_5-\gamma+1)(K_4-K_5)}{K_4 K_5}. \]
    \item From Lemma \ref{lemma:a2-perf},
    \[ (1+\delta')^{2L} \leq 1 + \delta, \]
    \[ \eps_4 = \left( 1-\frac{K_4}{K_3} \right) \cdot \min(\eps_3, \frac{2\delta'}{1+\delta'}). \]
    \item From Lemma \ref{lemma:a3-perf},
    \[ \left( 1-\frac{K_2}{(K_2-K_3)L}-\frac{K_2-K_3}{K_2} \right) \cdot \left( 1- \frac{K_2}{K_1} \left( 1-\frac{K_2}{(K_2-K_3)L} \right)^{-1} \right) \geq \frac{1}{10^5}. \]
    \item From Theorem \ref{thm:new-algo-perf},
    \[ \eps_5 \leq \frac{\eps_4}{10^5}. \]
    \item From Theorem \ref{thm:main-bifactor},
    \[ \eps_6 \leq \frac{\eps_5}{e^{\gamma_0}}. \]
    \item From Lemma \ref{lemma:char-func},
    \[ \eps_7 \leq \frac{K_1-\gamma+1}{2K_1-\gamma+2} \cdot \eps_5. \]
    \item From Lemma \ref{lemma:mu2-perf},
    \[ \eps_8 \leq \frac{\eps_7}{1000}. \]
\end{enumerate}

\end{document}